\documentclass[aps,prd,superscriptaddress,nofootinbib,amsmath,amsfonts,showkeys,showpacs,preprintnumbers,notitlepage,10pt,english]{revtex4-1}
\usepackage{amsmath}
\usepackage{amssymb}
\usepackage{babel}
\usepackage{graphicx}
\usepackage{dcolumn}
\usepackage{bm}
\usepackage[colorlinks,citecolor=blue,urlcolor=blue,linkcolor=blue]{hyperref}
\usepackage[figtopcap]{subfigure}
\usepackage{color}
\makeatletter

\usepackage[dvipsnames]{xcolor}

\usepackage{hyperref}
\usepackage{amsthm}
\theoremstyle{definition}

\theoremstyle{plain}

\allowdisplaybreaks

\@ifundefined{textcolor}{}{%
 \definecolor{BLACK}{gray}{0}
 \definecolor{WHITE}{gray}{1}
 \definecolor{RED}{rgb}{1,0,0}
 \definecolor{GREEN}{rgb}{0,1,0}
 \definecolor{BLUE}{rgb}{0,0,1}
 \definecolor{CYAN}{cmyk}{1,0,0,0}
 \definecolor{MAGENTA}{cmyk}{0,1,0,0}
 \definecolor{YELLOW}{cmyk}{0,0,1,0}
 }

\begin{document}
\title{Realistic compact stars in conformal teleparallel gravity}

\author{G.G.L. Nashed}%
\email{nashed@bue.edu.eg}
\affiliation{Centre for Theoretical Physics, The British University in Egypt, P.O. Box 43, El Sherouk City, Cairo 11837, Egypt}

\author{Kazuharu Bamba}%
\email{bamba@sss.fukushima-u.ac.jp}
\affiliation{Faculty of Symbiotic Systems Science,
Fukushima University, Fukushima 960-1296, Japan}

\date{\today}
\begin{abstract}
We explore an interior solution of a physically symmetric vierbein with two unknown functions in conformal teleparallel gravity. The field equations can be described in a closed system for a particular form of the metric potentials and an appropriate anisotropic function. As a result, we find a new set of configurations consistent with observed pulsars. In particular, the boundary conditions for the interior spacetime are used for the Schwarzschild spacetime to constrain the conformal field that with a unit value through the surface of a compact object. Furthermore, we apply the present model to the pulsar $4U1608-52$ with approximated radius $R= 9.52 \pm 0.15$ km and mass $M= 1.74 \pm 0.14\, M_{\circledcirc}$. To analyze the stability, we also study the causality conditions and the adiabatic index by assuming the Tolman-Oppenheimer-Volkov equation, adiabatic index and the static state. Moreover, the consistency of the model under consideration with other pulsars is investigated.


\end{abstract}

\maketitle
\newpage
\section{Introduction}\label{S1}

It is one of the most important challenges to explore gravity theories with conformal invariance \cite{tHooft:2010xlr,Mannheim:2011ds,tHooft:2014swy}. This was because conformal invariant theories are predicted to be renormalizable and thus a conceivable model of quantum gravity, and the conceivable way to reach the issues of dark energy and dark matter \cite{Mannheim:2005bfa}.
It is an important defect that in such a formulation, the Lagrangian includes higher derivative expressions, yielding a fourth-order theory with the Gauss-Ostrogradsky ghost. However, under specific situations, the ghost will vanish and the theories are expected to be consistent ones \cite{Maldacena:2011mk,Anastasiou:2016jix}.

Another method to overcome the problem of ghosts is to modify gravitational theories and consider the construction of torsional gravity, in which instead of the Riemannian geometry, the teleparallelism is used \cite{Cai:2015emx}. In the teleparallel spacetime, the connection is written based on the teleparallel constraints in which the curvature vanishes. In this regard, two different classes exist.

The first class is the one known as the symmetric teleparallel gravity theories constructed using symmetric teleparallel geometry. In this geometry, the connection is symmetric where the torsion is zero. Consequently, the geometry is prescribed by the spacetime non-metricity. This geometry was used first in \cite{Nester:1998mp} to construct the symmetric teleparallel equivalent of general relativity (STEGR).  Further, the geometry of STEGR was extended to different modified structures  \cite{Adak:2004uh,Adak:2005cd,BeltranJimenez:2017tkd,BeltranJimenez:2018vdo,Jarv:2018bgs}.

In the second class, the connection is completely fixed by extra constraints of the metric compatibility. Thus, this geometry is depicted by the spacetime torsion which allows the construction of completely torsion gravitational theories. A specific model of this class is the so-called \lq{}\lq{} teleparallel equivalent of general relativity (TEGR) \rq{}\rq{}, that is completely equivalent to general relativity \cite{Aldrovandi:2013wha,Krssak:2018ywd}. This theory, TEGR, was first used by Einstein attempting to unify electromagnetic and gravitational fields \cite{einstein1928riemannian,Sauer:2004hj}. Moreover, TEGR theory was used as an alternative construction of Einstein's GR in  \cite{hayashi1967extended,Nashed:2011fg,Nashed:2020kjh,Nashed:2010ocg,Shirafuji:1996im,Nashed:2007cu,cho1976einstein,hayashi1977gauge}. The TEGR theory was also employed in the frame of a metric-affine gravitational theory \cite{Hehl:1994ue,Obukhov:2002hy}. Furthermore, an alternative TEGR theory have been constructed, namely, $f(T)$ gravity \cite{Ferraro:2006jd,Ferraro:2008ey,Bengochea:2008gz,Linder:2010py,Chen:2010va,hayashi1979new,Nashed:2018efg,Nashed:2018cth,PhysRevD.24.3312,Kofinas:2014owa,Kofinas:2014daa,Bahamonde:2016kba,Harko:2014aja,Bahamonde:2017wwk,
Itin:2016nxk,Nashed:2016tbj,Hohmann:2017duq,Chen:2019ftv}, which became a familiar theory to discuss the cosmic accelerated expansions and investigate gravitational waves \cite{Bamba:2013ooa,Cai:2018rzd,Farrugia:2018gyz,Hohmann:2018jso,Nunes:2018evm}. { Moreover,  a non-metricity  conformal symmetric teleparallel gravity have been studied in \cite{Gakis:2019rdd}. Also, new anisotropic star solutions in mimetic gravity by imposing the Tolman-Finch-Skea metric have been discussed  in \cite{Nashed:2022yfc}.}


In this work, we concentrate on TEGR constructed in \cite{Maluf:2011kf,Maluf:2012yn}. There are many advantages for using TEGR theory. For instance, energy, momentum and angular momentum can be defined in a consistent manner \cite{maluf1994hamiltonian,Maluf:2006gu, Maluf:2005kn}. Also in the extend TEGR theory, the conformal invariance should be taken into account. The main feature of physical properties is that it is scale invariance, therefore, conformal invariance can be regarded as an extension of this invariance. The conformal term will have a the coordinate dependence. We here explain the meaning of a conformal transformation by consider the spacetime ($\cal{M}$, $g_{ab}$), with $\cal{M}$ a smooth $D$-dimensional manifold, $g_{ab}$ the metric in $\cal{M}$. We use the following conformal transformation:
\begin{equation}
\tilde{g}_{\mu \nu}=e^{2\zeta(x)}g_{\mu \nu}(x), \label{eq1}
\end{equation}\label{q1}
with $e^{2\xi(x)}$ being a smooth function of the coordinates. Therefore, the expression $e^{2\xi(x)}$ is the conformal factor. By this transformations, angles are kept and distances between two points prescribed by the same coordinate are alerted. The transformation (\ref{eq1}) becomes trivial when $\zeta(x)=\mathrm{const.}$ Any theory that respect transformation given by Eq. (\ref{eq1}) is called a conformal field theory in which the nature of physics has a scale invariance.

There are two explanations of the transformation (\ref{eq1}). One is for a constant background metric and the other is for a dynamical metric. For a constant metric, the transformation can be regarded as a physical symmetry of a dynamical system. On the other hand, for a dynamical metric, the transformation has a diffeomorphism, in other words, a gauge symmetry appears. For a review on the topic of  conformal field theory one can refer to  \cite{Gaberdiel:1999mc,Hawking:1973uf,Fujii:2003pa, 2004sgig.book.....C}. Recently, there are many papers including  conformal transformations have been studies  \cite{Chamseddine:2013kea,Chamseddine:2014vna,Nashed:2019cwg,Hawking:2000bb, Armillis:2013wya,Nashed:2018nll}, in the area of cosmology \cite{Buchbinder:1992rb,Buchbinder:1984giy};  in higher-order curvature terms  \cite{Moon:2010wq}, and in quantum gravity  \cite{tHooft:2011aa}. Our main purpose is to find a novel spherically symmetric interior solution in a conformal TEGR theory and study its relevant physics.

The organization of this paper is as follows. In Sec. \ref{S1}, we explain the conformal TEGR theory including mathematical definitions of tensors and field equations of gravitation. In Sec.~\ref{S2}, the field equations are applied to a four dimensional spherically symmetric vierbein field with non-diagonal components. We study the system of differential equations
with specific forms of the unknowns of the ansate and the anisotropic function and succeed to derive the remaining unknowns with three constants in Sec.~\ref{S2}. In Sec.~\ref{S4}, we present the necessary conditions that must be satisfied by real compact stellar objects. In Sec.~\ref{property} we show that the model under consideration is non-singular in the center of a star and derive the gradient of its energy-momentum components as well as its speed of sound. In Sec.~ \ref{data} we apply the junction conditions and derive the form of the constants in terms of the physical quantities numerically for the stellar $4U1608-52$, with  approximated radius  $R= 9.52 \pm 0.15$ km  and mass $M= 1.74 \pm 0.14\, M_{\circledcirc}$.  Moreover, in Sec.~ \ref{data} we show that the model bypass a real compact stellar graphically.  In Sec.~\ref{stability} we show that our model is stable using different methods, Tolman-Oppenheimer-Volkoff (TOV) equation, Adriatic index, and the static state. Also in Sec.~ \ref{data}, we show the consistency of our model with other pulsars. Finally, the section is devoted to the arguments of the astrophysical consequences.

\section{The conformal TEGR}\label{S1}
The TEGR geometry is described by \footnote{The Latin labels ${\it i, j, \cdots}$ refer to the coordinates of the tangent spacetime and the Greek indices $\alpha$, $\beta$, $\cdots$ denote the refer to the components of the (co-)frame.} $\{{\it M},~e_{i}\}$, where $\it M$ is a four-dimensional manifold and $e_{i}$ ($i=1,2,3,4$) is the  parallel vector on $\it M$.
The contra-variant derivative of the vierbein field reads
\begin{equation}\label{q1}
  D_{\mu} {e_i}^\nu:=\partial_{\mu}
{e_i}^\nu+{\Gamma^\nu}_{\lambda \mu} {e_i}^\lambda= 0\,,
\end{equation}
where the Weitzenb\"{o}ck connection is given by
\begin{equation}\label{q2}
{{\Gamma^\lambda}_{\mu \nu} := {e_i}^\lambda~ \partial_\nu e^{i}{_{\mu}}}\,,
\end{equation}
with $\partial_{\rho}=\frac{\partial}{\partial x^{\rho}}.$
In the TEGR theory, we can defined the  metric as:
\begin{equation}\label{q3}
 {g_{\mu \nu} :=  \eta_{i j} {e^i}_\mu {e^j}_\nu}\,,
\end{equation}
with the Minkowski metric $\eta_{i j}=(+,-,-,- )$.
The torsion ${\mathrm{T}^\alpha}_{\mu \nu}$ and
the contortion tensors $\mathrm{K}^{\mu \nu}{}_\alpha$ are defined as
\begin{eqnarray} \label{q33}
\nonumber {T^\alpha}_{\mu \nu}  & := &
{\Gamma^\alpha}_{\nu \mu}-{\Gamma^\alpha}_{\mu \nu} ={e_i}^\alpha
\left(\partial_\mu{e^i}_\nu-\partial_\nu{e^i}_\mu\right), \qquad
{K^{\mu \nu}}_\alpha  :=
-\frac{1}{2}\left({T^{\mu \nu}}_\alpha-{T^{\nu
\mu}}_\alpha-{T_\alpha}^{\mu \nu}\right)\,. \label{q4}
\end{eqnarray}
The contraction of the torsion yields
\begin{equation}\label{Tv}
{ T_\nu := {T^\mu}_{\mu \nu}}\,.
\end{equation}
The torsion scalar is
\begin{equation}\label{Tor_sc}
{T := {T^\alpha}_{\mu \nu} {S_\alpha}^{\mu \nu}}\,,
\end{equation}
where ${S_\alpha}^{\mu \nu}$ is the superpotential tensor whose first pair are  skew symmetry. The superpotential tensor is described by:
\begin{equation}\label{q5}
{ {S_\alpha}^{\mu \nu} := \frac{1}{2}\left({K^{\mu\nu}}_\alpha+\delta^\mu_\alpha{T^{\beta
\nu}}_\beta-\delta^\nu_\alpha{T^{\beta \mu}}_\beta\right)}.
\end{equation}
The Lagrangian is given by
\begin{equation}\label{q6}
\mathcal{L}({e^i}_\mu)_{g}=\frac{|e|T}{2\kappa}\,, \qquad \qquad \textrm {where} \qquad \qquad e=\sqrt{-g}\,.
\end{equation}
In this study, we use the relativistic units in which $c=G=1$, where $c$ is the speed of light and $G$ is the Newtonian constant\footnote{{In the following, we put the coupling of the gravitational constant as $\kappa=1$.}}.

In the conformal formalism, the Lagrangian of Eq.~(\ref{q6}) changes through the conformal transformation
\begin{equation}\label{q77}
\bar{e}_{a \mu}=e^{\zeta(x)} e_{a \mu},
\end{equation}
where $\xi(x)$ is an arbitrary function of the coordinate $x$ ~\cite{Maluf:2012yn}. Thus, Eq. (\ref{q6}) must be altered, so that Eq. (\ref{q6}) will not be changed by the conformal transformation (\ref{q77}).
Under the conformal transformation, the Lagrangian becomes \cite{Maluf:2012yn}
\begin{equation}\label{q7}
\mathcal{L}({e^i}_\mu, \xi)_{G}=2\kappa|e|\left[6g^{\mu \nu}\partial_\mu \xi\partial_\nu \xi-\xi^2T-4g^{\mu \nu}\xi(\partial_\nu \xi) T_\mu\right]+{\cal  L}_m\,,
\end{equation}
where $\xi$ is a scalar field and $\mathcal{L}_m$ is the Lagrangian of the matter. The Lagrangian (\ref{q7}) is invariant through
\[ \xi \rightarrow {\bar \xi}=e^{-\xi(x)} \xi.\]
Making the variation of the Lagrangian (\ref{q7}) in terms of $\xi$, we find~\cite{Maluf:2012yn}
\begin{equation}\label{q8}
I\equiv \partial_\mu(e g^{\mu \nu} \partial_\nu \xi)\xi-\frac{e}{6}\xi^2 R-\frac{e {\cal T}}{6}=0\,,
\end{equation}
with $R$ the scalar curvature given by $eR=2\partial_\nu(eT^\nu)-eT$, where $|e|=e=\sqrt{-g}=\det\left({e^a}_\mu\right)$ the determinant of the metric and $T$ being the scalar torsion defined from Eq. (\ref{Tor_sc}).
Here ${\cal T}$ is the trace of the energy momentum tensor defined as ${\cal T}=g_{\mu \nu}{\cal T}^{\mu \nu}$.

The variation of the Lagrangian~(\ref{q7})
in terms of the vierbein $e_{a \mu}$ yields~\cite{Maluf:2012yn}
\begin{eqnarray}\label{q9}
&&I^{a \nu}\equiv \partial_\alpha(e\xi^2 S^{a \nu \alpha})-e\xi^2(S^{b  \alpha \nu} T_{b \alpha}{}^a-\frac{1}{4}e^{a \nu} T)
 -\frac{3}{2}ee^{a \nu}g^{\beta \mu}\partial_\beta \xi \partial_\mu \xi+3ee^{a \mu}g^{\beta \nu}\partial_\beta \xi \partial_\mu \xi
 +ee^{a \nu}g^{\beta \mu}  T_\mu \xi \partial_\beta \xi\nonumber\\
& & -e\xi e^{a \beta}g^{\nu \mu} ( T_\mu  \partial_\beta \xi+T_\beta  \partial_\mu \xi)
 -eg^{\beta \mu}  \xi T^{\nu a}{}_\mu \partial_\beta \xi-\partial_\mu[e g^{\beta \nu} e^{a \mu} \xi \partial_\beta \xi]+\partial_\rho[e g^{\beta \rho} e^{a \nu} \xi \partial_\beta \xi]-e^a{}_\beta {\cal T}^{\beta \nu}=0\,.
\end{eqnarray}
Here, ${\cal T}_\alpha{^\beta}={\cal T}_G{_\alpha{^\beta}}+{\cal T}_S{_\alpha{^\beta}}$ with ${\cal T}_G{_\alpha{^\beta}}$ the energy-momentum tensor of anisotropic fluid, given by
\begin{equation}
{\cal T}_G^\mu{}_\nu=\left(p_{_{_\perp}}+\rho\right)u^\mu u_a+p_{_{_\perp}}\delta_a{}^\mu+(p_r-p_{_{_\perp}})\varepsilon_a \varepsilon^\mu\,,
\end{equation}
where $u^\mu=[1,0,0,0]$ is a time-like vector and $\varepsilon^\mu=[0,1,0,0]$ is the unit space-like vector along the radial direction, so that $u^\mu u_\mu=-1$ and $\varepsilon^\mu\varepsilon_\mu=1$. Moreover, $\rho$ denotes the energy density, $p_r$ and $p_{_{_\perp}}$ are the pressures of the radial and  tangential directions, respectively. Furthermore, ${\cal T}_S{_\alpha{^\beta}}$ is the energy-momentum tensor of the scalar field, represented as
\begin{equation}
{\cal T}_S^\beta{}_\nu=\partial^\beta \xi\partial_\nu \xi-\delta^\beta_\nu \left[g^{\alpha \alpha_1}\partial_\alpha \xi\partial_{\alpha_1} \xi-\frac{V(\xi)}{2}\right]\,.
\end{equation}
The above equations determine the conformal TEGR gravitational theory in 4-dimension space.

\section{Interior spherically symmetric solution }\label{S2}
We explore the metric with spherical symmetry metric
\begin{equation}
ds^2=-a^2(r) \,dt^2+\frac{dr^2}{b^2(r)}+r^2d\Omega^2\,, \qquad {\textrm where}\qquad  d\Omega^2=(d\theta^2+\sin^2\theta d\phi^2)\,,\label{met1}
\end{equation}
where $a(r)$ and $b(r)$ are functions of the radial coordinate $r$. Equation  (\ref{met1}) can be generated using the following covariant vierbein field ~\cite{Bahamonde:2019zea}
\begin{equation}
e^i{}_{\mu}=\left(
\begin{array}{cccc}
{a(r)} & 0 & 0 & 0 \\
0 & \displaystyle\frac{\cos (\phi ) \sin (\theta )} {b(r)} & r \cos (\phi ) \cos (\theta )  & -r \sin (\phi ) \sin (\theta )  \\
0 &\displaystyle\frac{\sin (\phi ) \sin (\theta )} {b(r)}  & r \sin (\phi ) \cos (\theta )  & r \cos (\phi ) \sin (\theta ) \\
0 & \displaystyle\frac{\cos (\theta )} {b(r)} & -r \sin (\theta ) & 0 \\
\end{array}
\right)\label{tet}\,.
\end{equation}
Moreover, Eq. (\ref{tet}) can be rewritten in a diagonal vierbein product and a local Lorentz transformation \cite{Nashed:2020kjh}.


Substituting Eq. (\ref{tet}) into Eq. (\ref{Tor_sc}), we find
\begin{align}\label{torsc}
  T={\frac { 2\left( b-1  \right)  \left(2\, a' b r -a  +a b   \right) }{a
  {r}^{2}}}\,.
\end{align}

Employing Eq. (\ref{torsc}) in the field equations (\ref{q8}) and  (\ref{q9}),  we get
\begin{eqnarray} \label{syd}
&&-1/2\,{\kappa}^{2}{r}^{2} \left( 2\,\rho
  -V  + \xi^{2}b^{2} \right)=\xi^{2} b^{2}- \xi^{2}+2\,{r}^{2}b
 \xi \xi' b' +4\,\xi\,r b^{2}\xi'  -{r}^{2}b^{2}{\xi'}^{2}+2\,
 \xi^{2}rb b'  +2\,{r}^{2} b^{2}\xi\xi''\,,\nonumber\\
&&\nonumber\\
&& 1/2\,{\kappa}^{2}a  {
r}^{2} \left( 2\,p_r + {\xi'}^{2} b^{2
}+V  \right)=2\,b^{2}{r}^{2}\xi \xi' a'  + \xi^{2}ab^{2}+4\,a
rb^{2}\xi
 \xi'  +3\,a  {r}^{2} b^{2}
{\xi'}^{2}+2\,\xi^{2}b^{2}ra'  - \xi^{2}a
\,,\nonumber\\
&&\nonumber\\
&&\frac {{\kappa}^{2}a r \left( 2\,p_t+V -
 {\xi'}^{2} b^{2} \right) }{2b  }=2\,\xi  \xi' r a' b  +2\,\xi \xi' a  rb'
  + \xi^{2}r
 a'b'
 +2\,\xi \xi' a b +
 \xi^{2}rb  a''  -a  b\,r{\xi'}^{2}+
\xi^{2}a b' +b{\xi'}^{2}a'  +2\,a  b r\xi\xi''
\,,\nonumber\\
&&\nonumber\\
&&\kappa^2r^2\,a(p_t-p_r-b^2{\xi'}^2)\equiv r^2\,a(\Delta(r)-\kappa^2\,b^2{\xi'}^2)
=2\,a {r}^{2}b \xi
 \xi' b' + \xi^{2}b
{r}^{2} a' b' - {\xi'}^{2}a  b^{2}-2\,a  r b^{2}\xi\xi' + \xi^{2} b^{2}{r}^{2}a''  -4\,a {r}^{2}b^{2}{\xi'}^{2
}\nonumber\\
&&\nonumber\\
&&+ \xi^{2}b  a
 b' r
- \xi^{2}b^{2}r\,a' +2\,a  {r}^{2}b^{2}\xi \xi''+ \xi^{2}a \,,\label{fec}
\end{eqnarray}
and the field equation of the scalar field takes the form:
\begin{eqnarray}
&&1/2{\kappa}^{2}a
  {r}^{2} \left( 2\,V - {\xi'}^{2} b^{2} \right)=
\xi  \left( \xi\,b[2 r
ba'  +2\,a  r b'  +a b+{
r}^{2} a'  b' + b{r}^{2} a''  -a /b]  +3\,\xi' r\,b[b{r} a' +a{r}^
{2}  b' +2\,a b +a  {r}b\xi''/\xi']  \right)\,,\nonumber\\
&&
\end{eqnarray}
where the prime $'$ means derivatives in terms of $r$. The expression of the energy-momentum tensor ${\cal T}^{\mu}_{\nu}$ is given by
\begin{align}\label{enmo}
  \mbox{diag}\,{\cal T}^\mu_{\nu}= \left(-\rho+\frac{V(r)}{2}-\frac{{\xi'}^2\,b^2}{2}\,,p_r+\frac{V(r)}{2}+\frac{{\xi'}^2\,b^2}{2}\,,p_t+\frac{V(r)}{2}+\frac{{\xi'}^2\,b^2}{2}\,,
  p_t+\frac{V(r)}{2}+\frac{{\xi'}^2\,b^2}{2}\right)\,,
\end{align}
where $V(r)$ is the potential field and $\xi$ is the conformal field.
$\Delta(r)$ in (\ref{syd}) is a parameter to express the anisotropy of compact objects. It is confirmed that Eqs. (\ref{syd}) are equivalent to those shown in \cite{Singh:2019ykp,Nashed:2020kjh,Nashed:2020buf} for $\xi(r)=V(r)=0$.

 The above system of differential equations consist of five independent equations in eight unknown functions, $a(r)$, $b(r)$, $\rho(r)$, $p_r(r)$, $p_t(r)$, $V(r)$, $\xi(r)$ and $\Delta(r)$. Therefore, three more conditions are necessary to make this system a closed system to be solved. As two conditions among them, the forms of $a(r)$ and $b(r)$ are supposed to be:
 \begin{align}\label{pot}
a(r) =a_1-a_0r^2\,,\qquad \qquad b(r) =1+a_0r^2\,
\end{align}
with $a_1$ a dimensionless constant and $a_0$ a one with the dimension of length. These constants will be determined from the junction condition to match the interior solution with the exterior one. Equation (\ref{pot}) indicates that when  $r=0$, $a(r)=a_1$ and $b(r)=1$, which are finite in the center of a star. Moreover, the derivative of the two functions $a$ and $b$ are finite in the center. The third condition is acquired by substituting Eq. (\ref{pot}) into the last equation of (\ref{syd}), i.e.,  $\Delta(r)$ which yields the form\footnote{{ From now on we are going to use the relativistic units in which $G=c=1$, namely, $\kappa=1$.}}
\begin{eqnarray}\label{d11}
&& \Delta(r)=\left(b^{2}{r}^{2}a''  - b^{2}
r\,a' +b  {r}^{2} a'b' +b  a b' r+a -a
 b^{2} \right) \xi^{2}+ \left( 2\,a  b  {r}^{2} \xi'
b'  +2\,a   b^{2}{r}^{2}\xi''
  -2\,a  b^{2}r\xi' \right) \xi -4\,a  b^{2}{r}^{2}{\xi'}^{2}+a  b^{2}{r}^{2}{\xi'}^{2}\,,\nonumber\\
  &&
\end{eqnarray}
that is a second order differential equation in terms of $\xi$ if
\begin{eqnarray}\label{d1}
&&0= \left( 2\,a  b  {r}^{2} \xi'
b'  +2\,a   b^{2}{r}^{2}\xi''
  -2\,a  b^{2}r\xi' \right) \xi -4\,a  b^{2}{r}^{2}{\xi'}^{2}+a  b^{2}{r}^{2}{\xi'}^{2}\,,
\end{eqnarray}
whose solution reads
\begin{eqnarray}\label{d2}
&&\xi={\frac {16{a_0}^{2}}{ \left( c_1\,\mathrm{S}+2{a_0}^2
 \right) ^{2}}}\,,\nonumber\\
 && \Longrightarrow \Delta(r)={r}^{4}{a_0}^{
2}a_1-4\,{r}^{4}{a_0}^{2}-5\,{r}^{6}{a_0}^{3}\,,
\end{eqnarray}
where $\mathrm{S}= \ln (1+a_0\,r^2)$.
Using Eq. (\ref{pot})  as well as Eq.  (\ref{d2}) in  Eqs. (\ref{syd}),  we get the unknowns $V(r)$, $\rho(r)$, $p_r(r)$ and $p_t(r)$ in the forms
\newpage
\begin{eqnarray}\label{sol}
&&\rho=128{a_0}^{5}\Bigg\{  6\,a_1\,c_1\,\mathrm{S} +5\,{a_0}^{2}{r}^{4}c_1 \,\mathrm{S} +6\,c_1\,\mathrm{S} +5\,{r}^{2}a_1\,a_0\,c_1\, \mathrm{S} +10\,c_1\,\mathrm{S} {r}^{2}a_0-12\,c_1\,{r}^{4}{a_0}^{2}-12\,c_1\,a_0\,{r}^{2}a_1-12\,c_1\,a_1-12\,c_1\,{r}^{2}a_0\nonumber\\
&&+20\,{a_0}^{3}{r}^{2}+12\,a_1\,{a_0}^{2}+
10\,{a_0}^{4}{r}^{4}+10\,{r}^{2}a_1 \,{a_0}^{3} +12\,{a_0}^{2}  \Bigg\}\Bigg\{ \mathrm{S}^{5}{c_1}^{5}{r}^{2}a_0-\mathrm{S}^{5}{c_1 }^{5}a_1-10\,\mathrm{S}^{4}{c_1}^{4}a_1\,{a_0}^{2} +10\,\mathrm{S} ^{4}{c_1}^{4}{r}^{2} {a_0}^{3}+32\,{r}^{2}{a_0}^{11}\nonumber\\
&&+40\, \mathrm{S}^{3}{c_1}^{3}{r}^{2}{a_0}^{5}-40\, \mathrm{S}^{3}{c_1}^{3}a_1\,{a_0}^{4}+80\,\mathrm{S}^{2}{c_1}^{2}{r}^{2}{a_0}^{7}-80\, \mathrm{S}^{2}{c_1}^{2}a_1\,{a_0}^{6}+80\,\mathrm{S}  c_1 \,{r}^{2}{a_0}^{9}-80\,\mathrm{S}  c_1\,{a_0}^{8}a_1-32\,a_1\,{a_0}^{10}\Bigg\}^{-1}
\,, \nonumber\\
&&p_r=-128{a_0}^{5}\Bigg\{ 2\,c_1\,\mathrm{S} {r}^{2}a_0+5\,{a_0}^{2}{r} ^{4}c_1\,\mathrm{S} -2\,c_1\, \mathrm{S} -3\,{r}^{2}a_1\,a_0\,c_1\,\mathrm{S}  +4\,c_1\,a_0\,{r}^{2}a_1-2\,a_1\,c_1\,\mathrm{S}+4\,c_1\,a_1+4\,c_1\,{r}^{2}a_0+4\,c_1\,{r}^{4}{a_0}^{2}\nonumber\\
&&+4\,{a_0}^{3}{r}^{2} -6\,{ r}^{2}a_1\,{a_0}^{3}-4\,{a_0}^{2}+10\,{a_0}^{4}{r} ^{4}-4\,a_1\,{a_0}^{2}  \Bigg\}\Bigg\{ \mathrm{S}^{5}{c_1}^{5}{r}^{2} a_0-\mathrm{S}^{5}{c_1}^{5}a_1-10\, \mathrm{S}^{4}{c_1}^{4}a_1\,{a_0}^{2}+10\,\mathrm{S}^{4}{c_1 }^{4}{r}^{2}{a_0}^{3}-32\,a_1\,{a_0}^{10}\nonumber\\
&&+40\,\mathrm{S}^{3}{c_1}^{3}{r}^{2}{a_0}^{5}-40\,\mathrm{S}^{3}{c_1 }^{3}a_1\,{a_0}^{4}+80\,\mathrm{S}^{2}{c_1}^{2}{r}^{2}{a_0}^{7}-80\,\mathrm{S}^{2}{c_1 }^{2}a_1\,{a_0}^{6}+80\,\mathrm{S} c_1\,{r}^{2}{a_0}^{9}-80\,\mathrm{S} c_1\,{a_0}^{8}a_1+32\,{r}^{2}{a_0}^{11}\Bigg\}^{-1}\,,\nonumber \\
&&p_t=128{a_0}^{5}\,\Bigg\{  6\,c_1\,\mathrm{S}{r}^{2}a_0+5\,{a_0}^{2}{r}^{4}c_1\,\mathrm{S}+2\,c_1\,\mathrm{S} +{r}^{2}a_1\,a_0\,c_1\,\mathrm{S} -4\,c_1\,a_0\,{r}^{2}a_1+2\,a_1\,c_1\,\mathrm{S}-4\,c_1\,a_1-4\,c_1\,{r}^{2}a_0-4\,c_1\,{r}^{4}{ a_0}^{2}\nonumber \\
&& +12\,{a_0}^{3}{r}^{2}+2\,{r}^{2}a_1\,{a_0} ^{3}+4\,{a_0}^{2}+10\,{a_0}^{4}{r}^{4}+4\,a_1\,{a_0}^{2}\Bigg\}\Bigg\{\mathrm{S}^{5}{c_1}^{5}{r}^{2}a_0- \mathrm{S}^{5}{c_1}^{5}a_1-10\,\mathrm{S}^{4}{c_1}^{4}a_1\,{a_0}^{2}+10\, \mathrm{S}^{4}{c_1}^{4}{r}^{2}{a_0}^{3}+32\,{r}^{2}{a_0}^{11}\nonumber \\
&&+ 40\,\mathrm{S}^{3}{c_1}^{3}{r}^{2}{a_0}^{5}-40\,\mathrm{S}^{3}{c_1}^{3}a_1\,{a_0}^{4}+80\,\mathrm{S}^{2}{c_1 }^{2}{r}^{2}{a_0}^{7}-32\,a_1\,{a_0}^{10}-80\,\mathrm{S}^{2}{c_1}^{2}a_1\,{a_0}^{6}+80\, \mathrm{S} c_1\,{r}^{2}{a_0}^{9 }-80\,\mathrm{S} c_1\,{a_0}^{8} a_1\Bigg\}^{-1}\, , \nonumber\\
&& V(r)=\frac{256{a_0}^{5}}{\mathrm{S}}\Bigg\{24\,{a_0}^{4}-6 \mathrm{S}^{2}{c_1}^{2}a_1+6\mathrm{S}^{2}{c_1}^{2}-20 {r}^{2}a_1{a_0}^{3}c_1\,\mathrm{S}-20{r}^{2}a_1\,{a_0}^{5} +36{r}^{2}a_0{c_1}^{2}a_1\mathrm{S} -5{r}^{2}a_1a_0{c_1}^{2}\mathrm{S}^ {2} +80{a_0}^{2}{r}^{4}{c_1}^{2}\nonumber\\
&&+36\,{c_1}^{2}a_1\,\mathrm{S}+72\,{a_0}^{2}c_1\,a_1-120\,{r}^{4}{a_0}^{4}c_1+24\,c_1\,\mathrm{S} {a_0}^{2}-120\,{r}^{2}{a_0}^{3}c_1+88\,{r}^{ 2}{a_0}^{5}+60\,{a_0}^{6}{r}^{4}-24\,a_1\,{a_0}^{4 }+60\,{a_0}^{4}{r}^{4}c_1\,\mathrm{S}\nonumber\\
&& -24a_1c_1\mathrm{S} {a_0}^{2}-60{r}^{4}{a_0}^{2}{c_1}^{2}\mathrm{S} +22{r}^{2}a_0{c_1}^{ 2}\mathrm{S}^{2}+88{r}^ {2}{a_0}^{3}c_1\mathrm{S} -60 {r}^{2}a_0{c_1}^{2}\mathrm{S} -80a_1\,{c_1}^{2}{r}^{2}a_0+15{a_0} ^{2}{r}^{4}{c_1}^{2}\mathrm{S}^{2}+72{r}^{2}{a_0}^{3}c_1a_1\Bigg\}  \nonumber\\
&&\times\Bigg\{ {c_1}^{6} \mathrm{S}^{6}+12\,{c_1}^{5}\mathrm{S}^{5}{a_0}^{2}+60\, {c_1}^{4}\mathrm{S}^{4}{a_0}^{4}+160\,{c_1}^{3}\mathrm{S}^{3}{a_0}^{6}+240\,{c_1} ^{2}\mathrm{S}^{2}{a_0}^{8}+192\,c_1\,\mathrm{S} {a_0}^{10}+64\,{a_0}^{12} \Bigg\}^{-1}\,.
\end{eqnarray}
It is important to note that the anisotropic force is given by $\frac{2\Delta}{r}$, which is repulsive if $p_r-p_t<0$, while it is attractive when $p_r-p_t>0$. The mass of a sphere with its radius $r$ is expressed by \cite{Lasky:2008fs,Fattoyev:2019aht}
\begin{align}\label{mas}
M(r)={\int_0}^r \left[4\pi \rho(\zeta) -\frac{{\xi'(\zeta)}^2b(\zeta)}{2}+\frac{V(\zeta)}{2}\right]\zeta^2 d\zeta\,.\end{align}
Combining Eq. (\ref{sol}) with Eq. (\ref{mas}), we have  a very length  form of $M(r)$ which has the following asymptotic form\footnote{The numerical values of the constants $a_0$, $a_1$, and $c_1$ that will be fixed from the junction condition in Sec.~\ref{S4} using the pulsar  $4U1608-52$, whose approximated radius  $R= 9.52 \pm 0.15$ km  and mass $M= 1.74 \pm 0.14\, M_{\circledcirc}$, are used in Eq. (\ref{mas}).}:
\begin{align}\label{mas1}
M(r)\approx-6.358950030\times10^{11}r+3.712536047\times10^{12} r^3-2.148986652\times10^{13}r^5\,,\end{align}

With Eq. (\ref{mas1}), we find a parameter describing the compactness of a spherically symmetric star \cite{Singh:2019ykp}
\begin{eqnarray}\label{gm1}
&&u(r)\approx\frac{2M(r)}{r}=-1.271790006\times10^{12}+7.425072094\times10^{12} r^2-4.297973304\times10^{13} r^4.
\end{eqnarray}

In the next section, we investigate the conditions, with which a realistic star must satisfy and test whether Eq. (\ref{sol}) can be met.

\section{Necessary conditions of a realistic star}\label{S4}
Necessary conditions that a realistic star must satisfy are summarized as follows.
\vspace{0.1cm}\\
$\bigstar$
The metric potentials $a(r )$ and $b(r )$ and the matter components
 $\rho(r)$, $p_r(r)$, $p_t(r)$  have finite values in the center of a star and are regular and free from singularity. \vspace{0.1cm}\\
$\bigstar$
The matter density $\rho$ should be positive in all regions of a star, $0\leq \rho(r)$. Moreover, the value of $\rho$ should be positive in the center and finite and it decreases across the boundary of the interior of a star, namely, $0\geq\rho'(r)=\frac{d\rho}{dr}$.\vspace{0.1cm}\\
$\bigstar$
The radial pressure $p_r$ as well as the tangential one $p_t$
should be positive within a star, $p_r\geq0$, $p_t\geq0$.
Moreover, the derivative of the pressure have to be
negative within a star, $\frac{dp_r}{dr}< 0$, $\frac{dp_t}{dr}< 0$.
In the center of a star, the anisotropy vanish, namely,
$\Delta(r\rightarrow 0) = 0$.\vspace{0.1cm}\\
$\bigstar$
The requirement of the energy conditions is necessary for anisotropic fluid sphere. These conditions are described as the following forms, that is satisfied in all points of a star \vspace{0.1cm}\\
(i) The dominant energy conditions (DEC): $\rho\geq \lvert p_r\lvert$ and
$\rho\geq \lvert p_t\lvert$.\vspace{0.1cm}\\
(ii) The strong energy condition (SEC): $p_r+\rho > 0$, $p_t+\rho > 0$, $\rho-p_r-2p_t > 0$.\vspace{0.1cm}\\
(iii) The weak energy condition (WEC): $p_r+\rho > 0$, $\rho> 0$.\vspace{0.1cm}\\
(iv) The null energy condition (NEC): $p_t+\rho > 0$, $\rho> 0$.\vspace{0.1cm}\\
$\bigstar$
The causality condition for a realistic star, that is, the sound speed must be smaller than unity (in the relativistic units of $c = G = 1$) in the interior of
the star, that is,  $1\geq \frac{dp_t}{dr}\geq 0$ and $1\geq\frac{dp_r}{dr}\geq 0$.\vspace{0.1cm}\\
$\bigstar$
The interior solution have to match with
the exterior Schwarzschild one in the star surface for $\xi=1$ \cite{Nashed:2019cwg,Nashed:2018nll}.\vspace{0.1cm}\\
$\bigstar$
The adiabatic index have to be larger than $\frac{4}{3}$ to have a stable model.\vspace{0.1cm}\\
$\bigstar$
For the stability analysis of an anisotropic star,
it has been suggested \cite{HERRERA1992206} that the relation $0<v_t{}^2-v_r{}^2<1$ must be met. Here, $v_t$ and $v_r$ are the transverse and radial speeds, respectively.

By using the physical conditions, we study whether the present model can satisfy the relations mentioned above.

\section{The physical conditions of the present model}\label{property}
We examine Eq. (\ref{sol}) and confirm that it is appropriate for a realistic star.

\subsection{A free singular model}
First, the following conditions are satisfied by the metric potentials
\begin{align}\label{sing}
a(r\rightarrow 0)=a_1\,\qquad  \textrm{and} \qquad b(r\rightarrow 0)=1.
\end{align}
If this condition is met, the metric potentials have finite value in the center of a star. Additionally, the gradients of the metric potentials have finite value in the center such as $a'_{r\rightarrow0}=b'_{r\rightarrow0}=0$. This  means that the metric potentials are not singular in the center.\vspace{0.1cm}\\

Second, the components of the fluid given by Eq. (\ref{sol}) in the center of a star are express as
\begin{align}\label{reg}
\rho(r\rightarrow 0)=\frac{48a_1c_1-48a_0{}^2-48a_1a_0{}^2}{a_1a_0{}^5}\,, \qquad p_r(r\rightarrow 0)=p_t(r\rightarrow 0)=\frac{16a_1c_1-16a_0{}^2-16a_1a_0{}^2}{a_1a_0{}^5}.
\end{align}
The above equations show that the energy density is positive if $\frac{48a_1c_1-48a_0{}^2-48a_1a_0{}^2}{a_1a_0{}^5}>0$, and that the spacetime becomes isotropic in the center. Moreover, the radial and tangential pressures become positive value when $\frac{16a_1c_1-16a_0{}^2-16a_1a_0{}^2}{a_1a_0{}^5}>0$, otherwise they become negative.
Furthermore, in Ref.~\cite{1971reas.book.....Z}, radial pressure should equal or to or weaker than density in the center, $\frac{p_r(0)}{\rho(0)}\leq 1$. With the constrains $\frac{p_r(0)}{\rho(0)}$ given by Eq. (\ref{reg}), it is seen that this condition can be satisfied.\\

Third, we describe the gradient of energy density, radial and tangential
pressures in Appendix {\color{blue} (A)}. In the next section, we consider these quantities by fixing the numerical values of $a_0$, $a_1$, and $c_1$.\\

Fourth, the radial and transverse sound speeds ($c = 1$) are presented in Appendix {\color{blue} (B)}. We show these components for the concrete values of $a_0$, $a_1$, and $c_1$.

\subsection{Junction conditions}
It is supposed that the exterior region of a non-rotating star is vacuum and
is expressed by the Schwarzschild metric, given by \footnote{It is shown that the general solution of a non-charged spherically symmetric spacetime in conformal teleparallel theory is just the Schwarzschild solution \cite{Nashed:2018nll}.}:
\begin{eqnarray}\label{Eq1}  ds^2= -\left(1-\frac{2M}{r}\right)dt^2+\left(1-\frac{2M}{r}\right)^{-1}dr^2+r^2d\Omega^2,
 \end{eqnarray}
with $M$ the total mass.
We have to  match

The interior metric (\ref{met1}) must be matched with
the exterior Schwarzschild one (\ref{Eq1}) in the star
boundary $r = l$. By using the continuity of the metric at $r = l$, we find
\begin{eqnarray}\label{Eq2}
 a(r\rightarrow l)=\left(1-\frac{2M}{l}\right)\,,
 \qquad \qquad b(r\rightarrow l)=\left(1-\frac{2M}{l}\right)\,, \qquad \qquad \xi(r\rightarrow l)=1\,.
 \end{eqnarray}
With these conditions, the constants $a_0$, $a_1$ and $c_1$ are determined as
 \begin{eqnarray}\label{Eq2}
 a_0=-\frac{2M}{l^3}\,, \qquad  a_1=-\frac{4M-l}{l}\,, \qquad c_1=-\frac{8M(M\pm l^3)}{l^6\ln\biggl(\frac{l-2M}{l}\biggr)}\,.
\end{eqnarray}
We use the numerical values of these constants when we deal with observations.

 \section{Comparison of theoretical model with the observations of real stars} \label{data}

{ Now, we convert back to the standard  physical units to accurately determine the values of the model parameters under consideration numerically. Through the use of the masses and the radii of observed pulsars with
the physical conditions we referred above,
one can fix the constant parameters $a_0$ and $a_1$ and check its viability. Now, we employ the observational
limits of the pulsar \textrm{4U 1608-52} , whose mass
$M = 1.74\pm 0.14 M_\circledcirc$ and $R = 9.52\pm0.15$ km, where $M (= 1.989×1030 kg)$ denotes the solar mass.
Then, the boundary conditions given by Eq. \eqref{Eq2} are able to
fix the dimensionless constants.}

We adopt the conditions listed previously to compare the present model to observational data. We study the pulsar \textrm{4U 1608-52} whose mass and radius are $M = 1.74\pm 0.14 M_\circledcirc$ and $R = 9.52\pm0.15$ km, respectively \cite{Gangopadhyay:2013gha}, and therefore we take $M=1.78 M_\circledcirc$ and  $R=9.67$km. We use the boundary conditions to examine the values of the constants $a_0=-0.005826829242$, $a_1=-0.08972078594$, and $c_1=-0.02952344694$. With these constants, we depict physical quantities.
We examine whether the present model can fit the observational data of realistic stars.

\begin{figure}
\centering
\subfigure[~Metric potentials]{\label{fig:pot1}\includegraphics[scale=0.3]{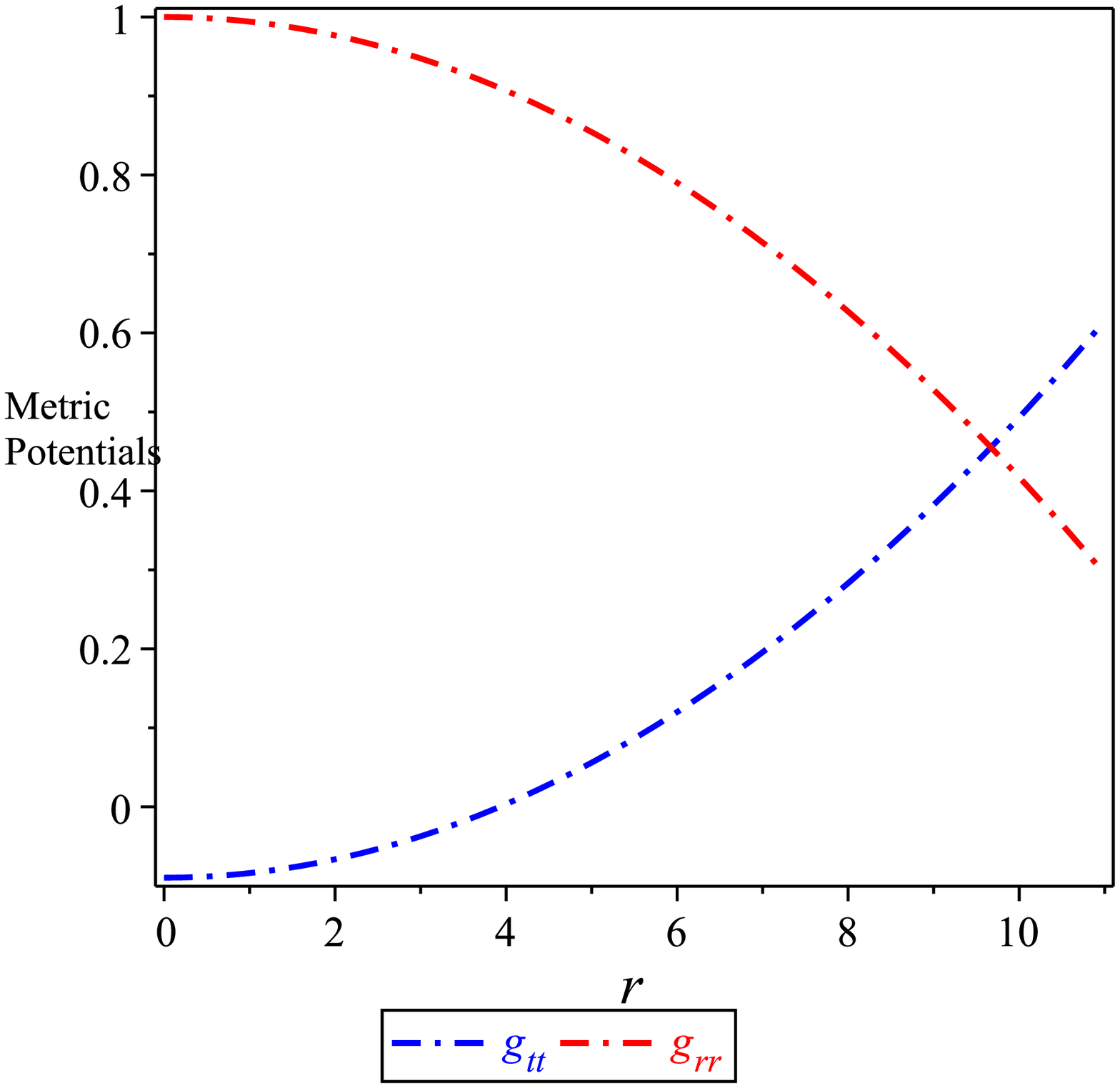}}
\subfigure[~Energy-density]{\label{fig:den}\includegraphics[scale=.3]{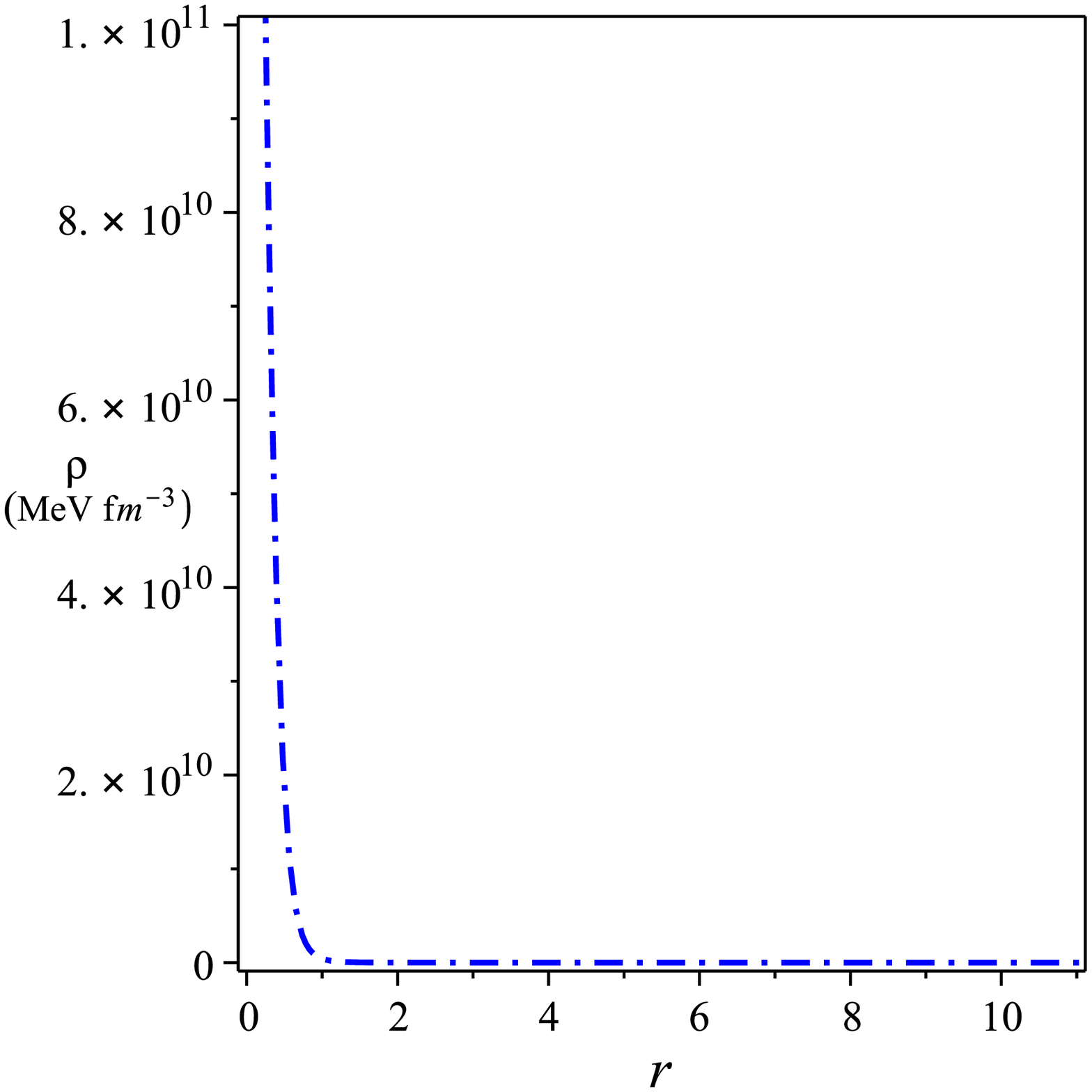}}
\subfigure[~Radial pressure]{\label{fig:pr}\includegraphics[scale=.3]{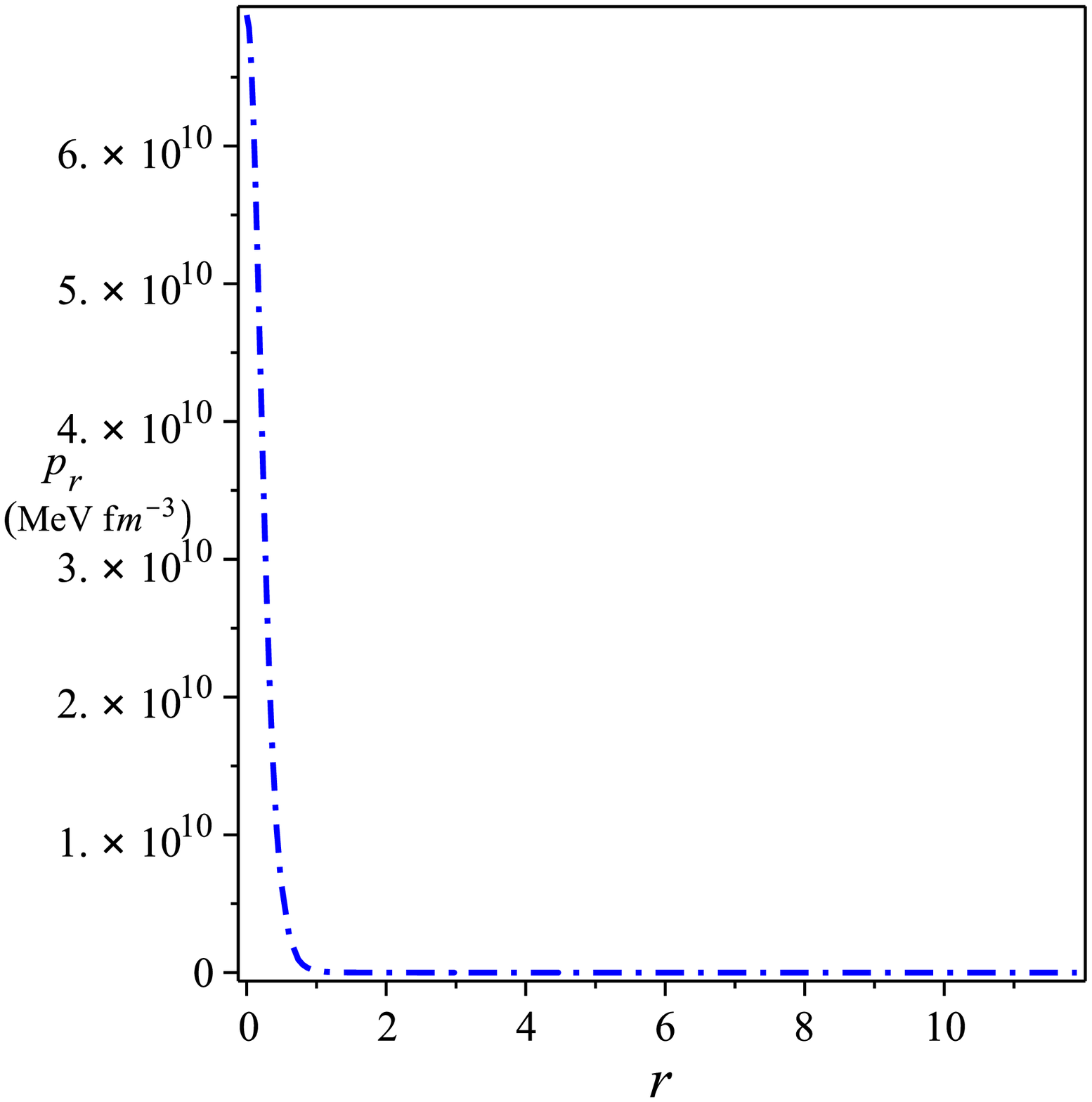}}
\subfigure[~Tangential pressure]{\label{fig:pt}\includegraphics[scale=.3]{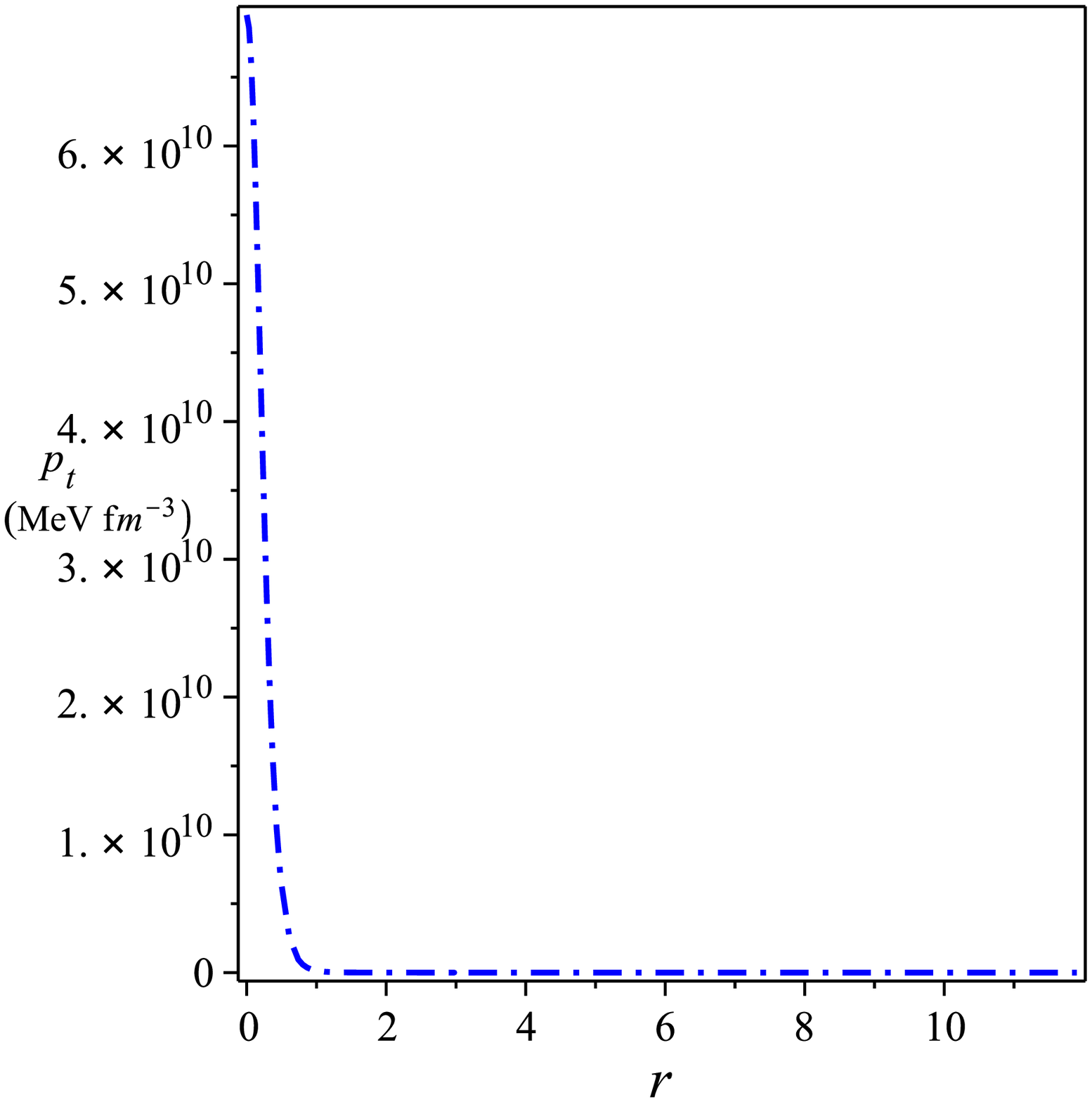}}
\caption[figtopcap]{\small{{Schematic plots of ~\subref{fig:pot1} the  metric potentials; ~\subref{fig:den} the energy-density; ~\subref{fig:pr} the radial  pressure and ~\subref{fig:pt} the tangential pressure. All of these plots are depicted  vs. the radial coordinate $r$.}}}
\label{Fig:1}
\end{figure}

Figure \ref{Fig:1} \subref{fig:pot1} shows the metric potentials for
\textrm{4U 1608-52}.
From this figure, we see that $a(r\rightarrow 0)=-0.08972078594$, and $b(r\rightarrow 0)=1$, both of which are finite in the center of a star.

Figures \ref{Fig:1} \subref{fig:den}, \ref{Fig:1} \subref{fig:pr} and \ref{Fig:1} \subref{fig:pt} plot the density, radial and tangential pressures, respectively. We see that the density, radial and tangential pressures are positive and they become large in the center and become smaller as the radii becomes larger. These behaviors are consistent with realistic stars.
\begin{figure}
\centering
\subfigure[~Anisotropy and anisotropic force]{\label{fig:An}\includegraphics[scale=0.3]{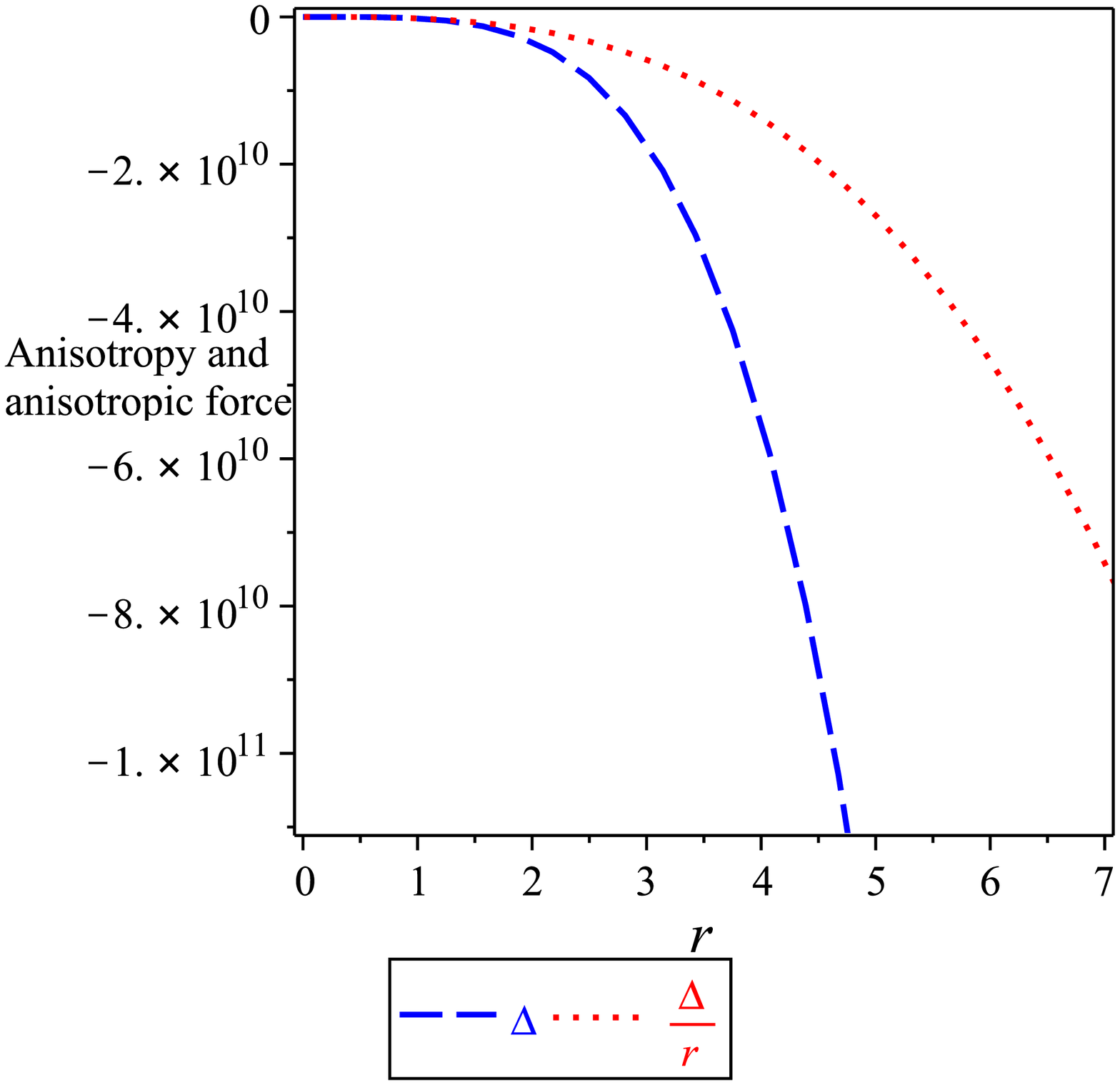}}
\subfigure[~Gradient of the energy-momentum components]{\label{fig:grad}\includegraphics[scale=.3]{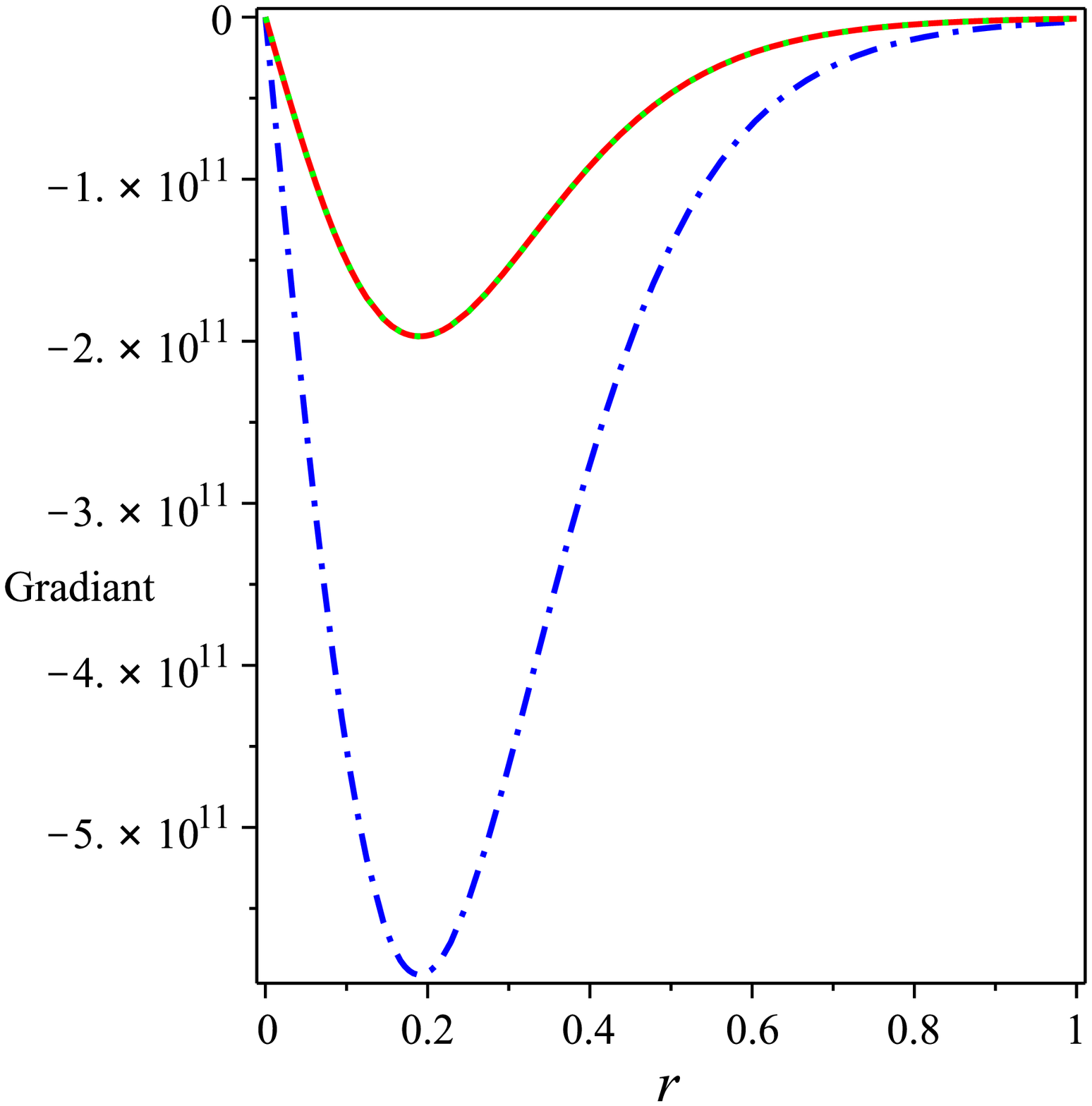}}
\subfigure[~Speed of sounds]{\label{fig:speed}\includegraphics[scale=.3]{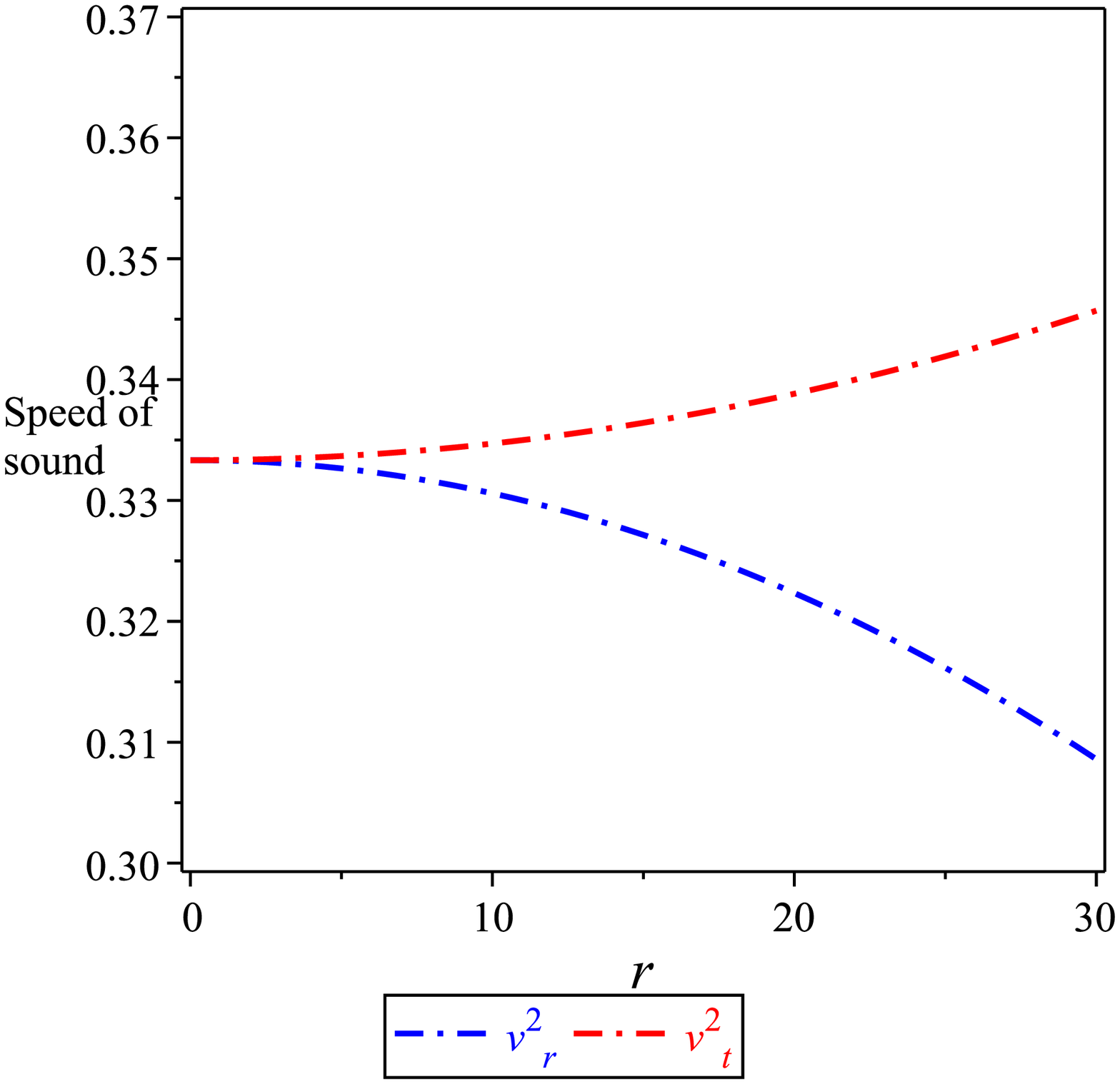}}
\subfigure[~Difference between the tangential and radial speed of sounds]{\label{fig:dspeed}\includegraphics[scale=.3]{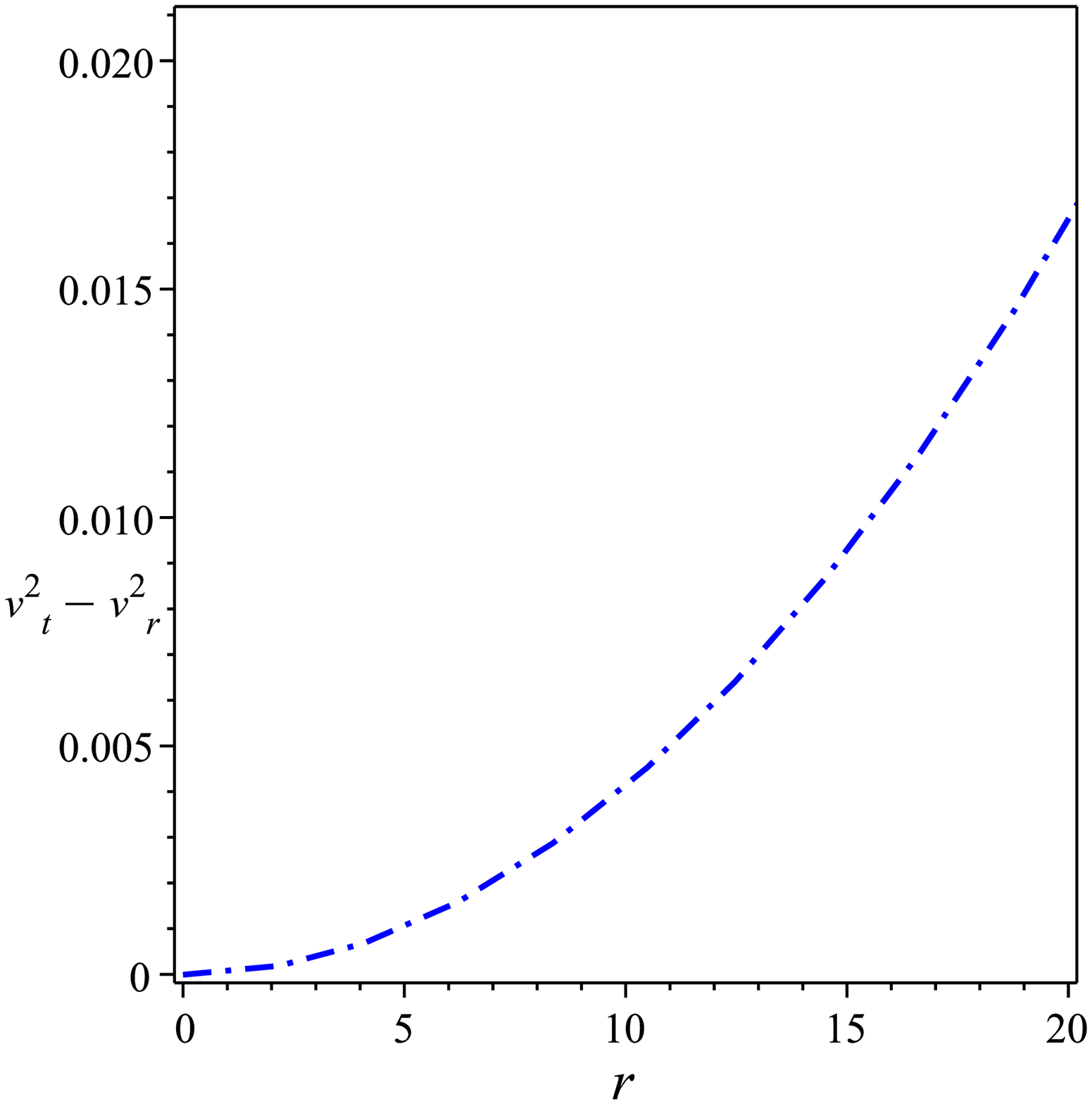}}
\caption[figtopcap]{\small{{Schematic plots of ~\subref{fig:An} the anisotropy and anisotropic force; ~\subref{fig:grad} the gradient of the energy-momentum components; ~\subref{fig:speed}  the speed of sound;  and ~\subref{fig:dspeed} the difference between the tangential and radial speed of sounds. All of the figures are depicted  vs. the radial coordinate $r$.}}}
\label{Fig:2}
\end{figure}
From figure \ref{Fig:2} \subref{fig:An}, we understand that the spacetime becomes isotropic in the center of a star and the anisotropy is smaller in the surface of a star. Specifically, we find the negative anisotropic force $\frac{\Delta}{r}$, that is, t is an attractive one due to the fact that $p_t-p_r<0$.
Figure \ref{Fig:2}  \subref{fig:grad} means that the gradients of the energy-momentum components, density, radial and tangential pressures presented in Appendix {\color{blue} (A)}, are negative. This ensures the decreasing of all of them as a whole of a star.
In  Figure \ref{Fig:2} \subref{fig:speed}, we depict the radial and tangential speeds  of sound, presented in Appendix {\color{blue} (B)}. These values are positive and less than unity, so that the causality condition can be met within a star. Figure \ref{Fig:2}  \subref{fig:dspeed} shows the difference between the tangential and radial speed of sounds that has a positive value as required for any realistic stellar.
\begin{figure}
\centering
\subfigure[~Weak  energy conditions]{\label{fig:WEC}\includegraphics[scale=0.3]{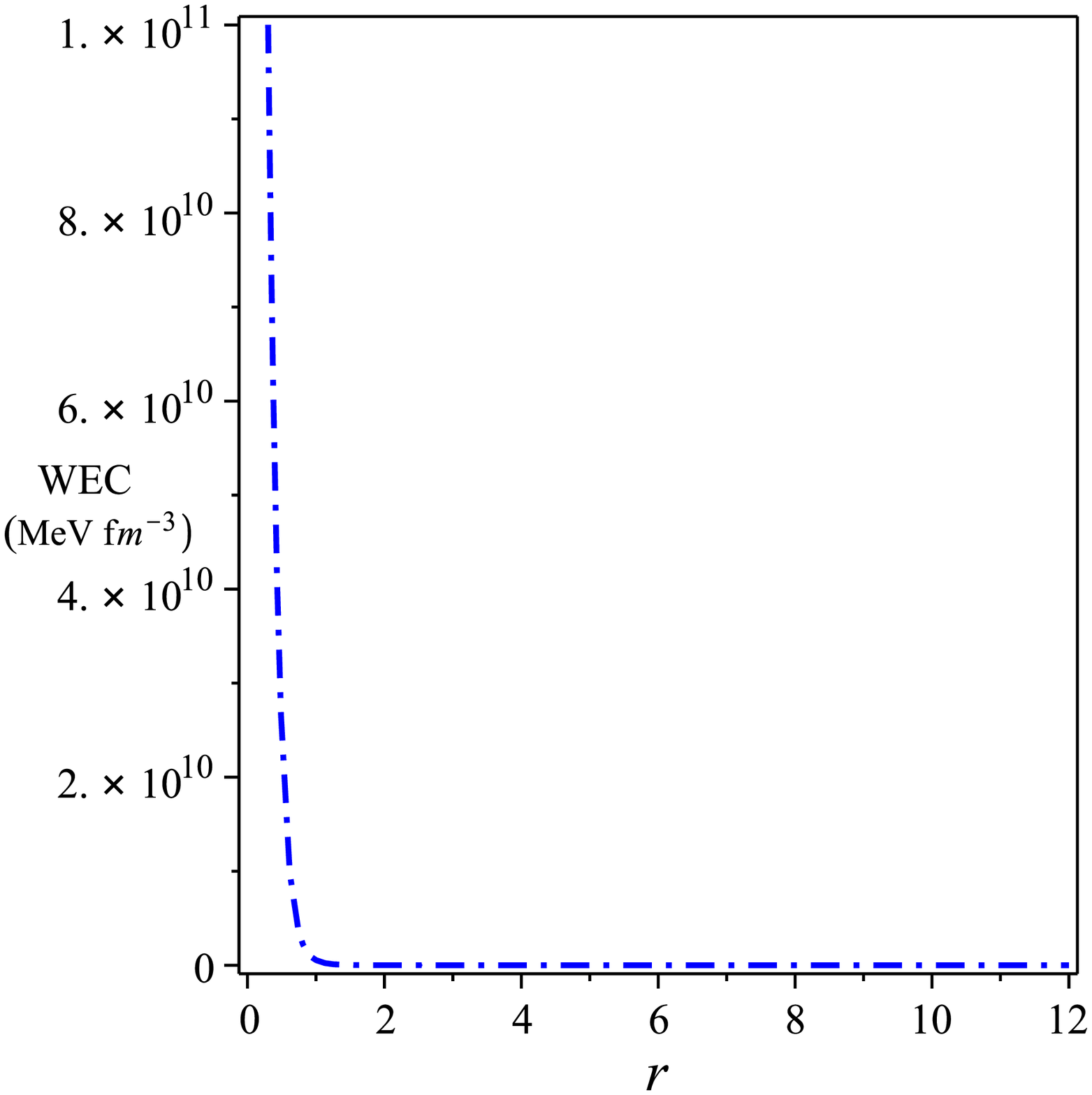}}
\subfigure[~Null  energy conditions]{\label{fig:NEC}\includegraphics[scale=0.3]{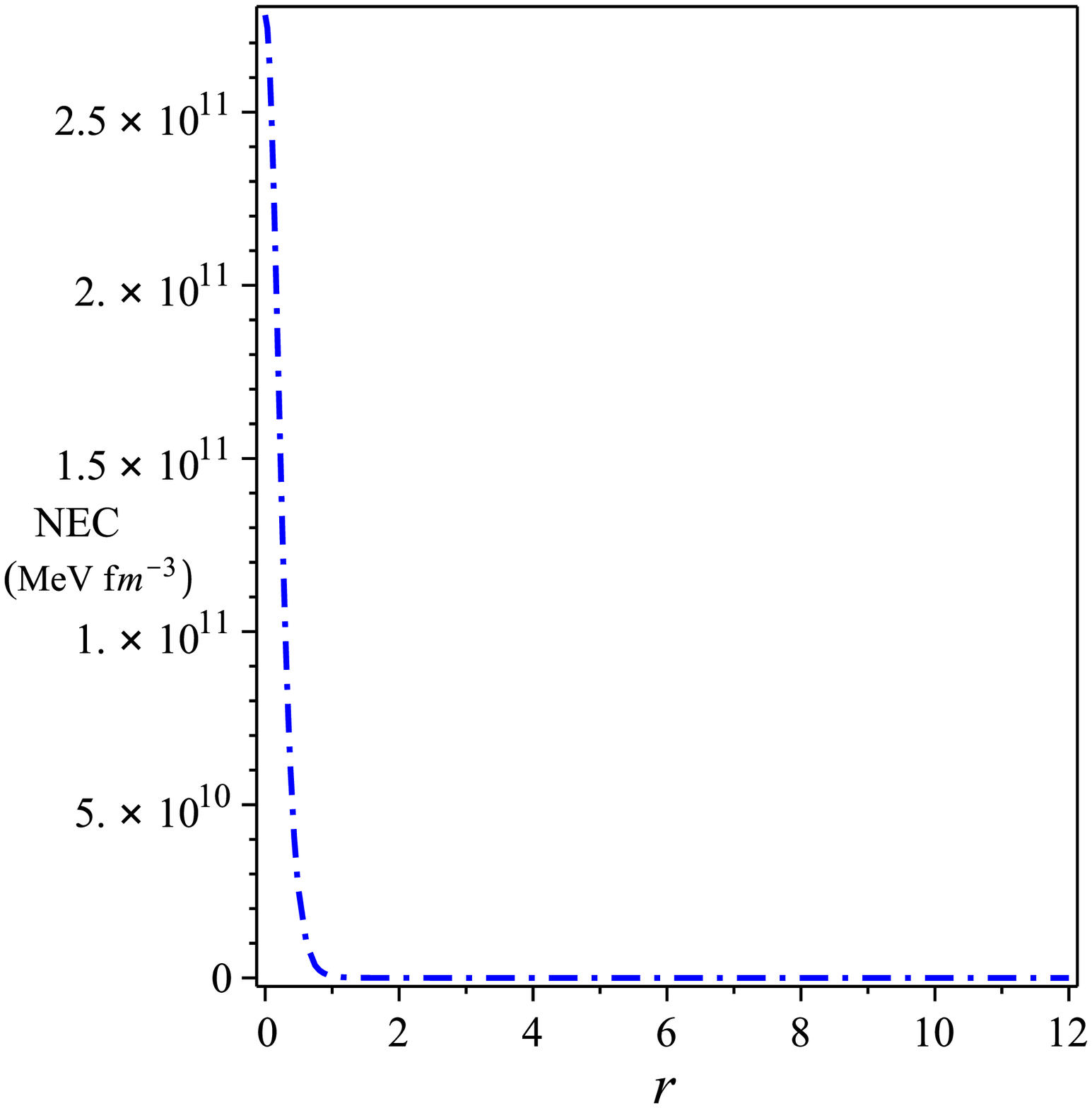}}
\subfigure[~Strong energy condition]{\label{fig:SEC}\includegraphics[scale=.3]{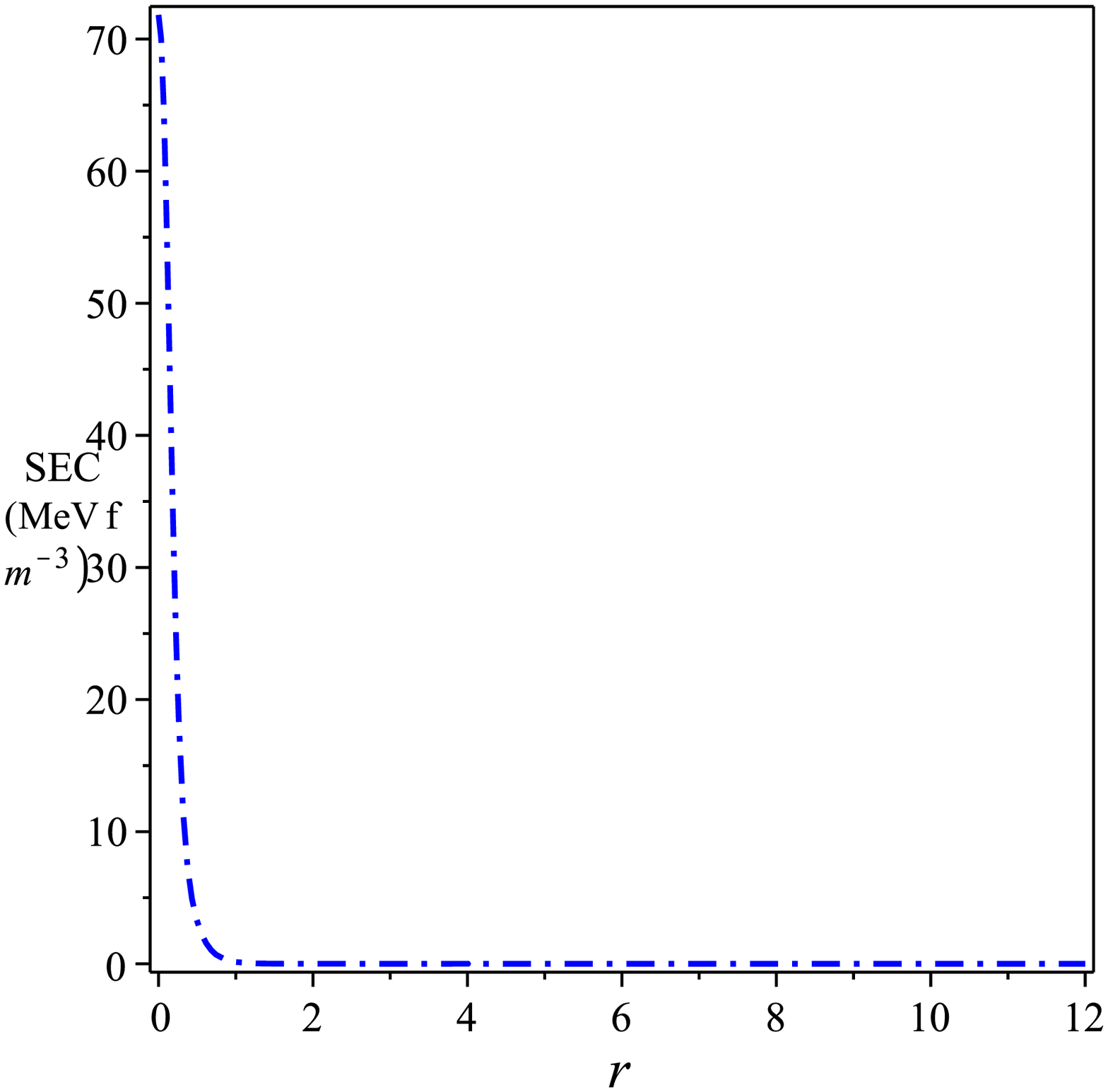}}
\subfigure[~Dominant energy condition]{\label{fig:DEC}\includegraphics[scale=.3]{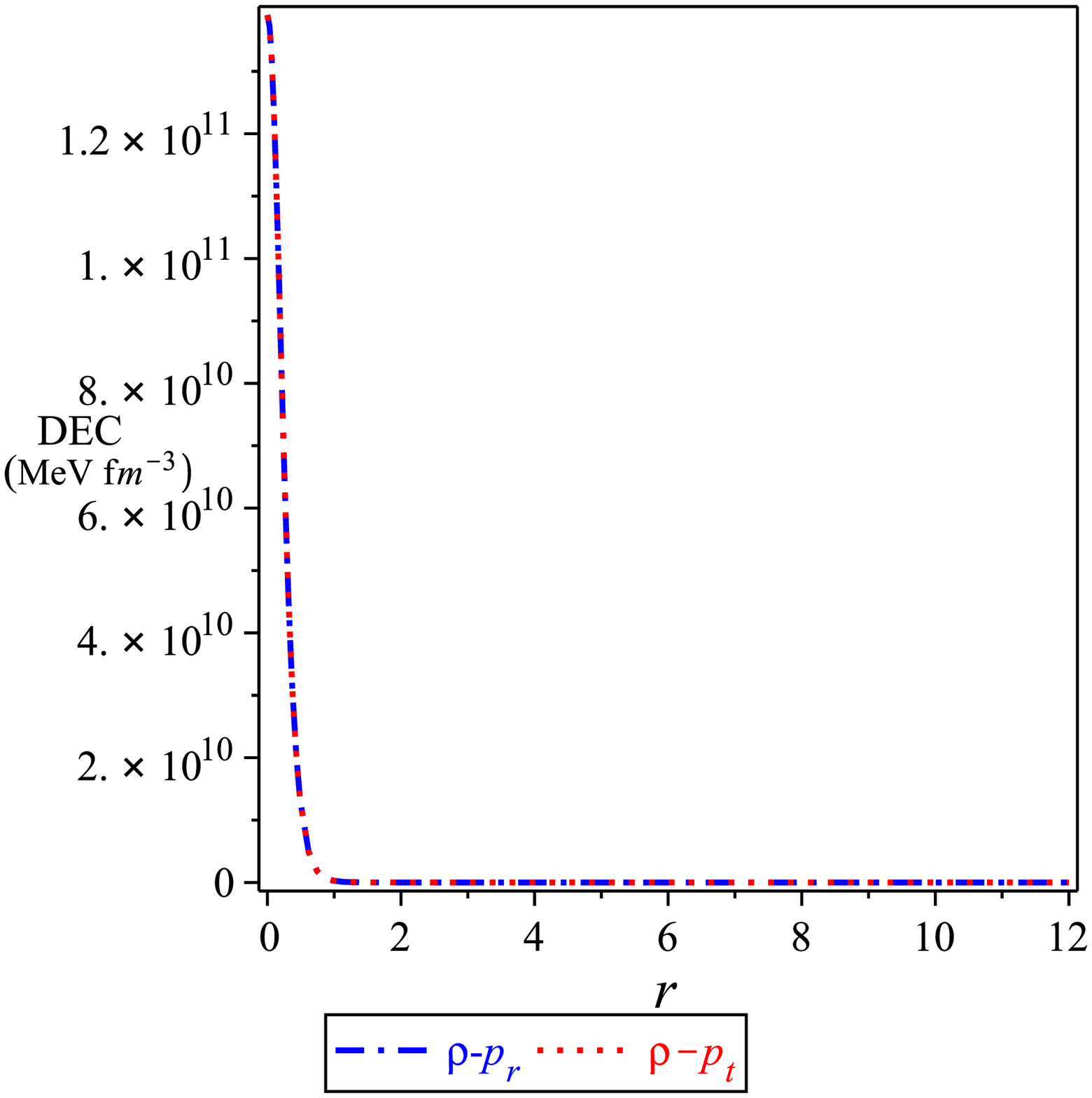}}
\caption[figtopcap]{\small{{Plot of the  ~\subref{fig:WEC}  weak, ~\subref{fig:NEC}  null, ~\subref{fig:SEC} strong and ~\subref{fig:DEC}  dominant energy conditions given by  Eq. (\ref{sol}). All these plots are depicted   vs. the radial coordinate  $r$. }}}
\label{Fig:3}
\end{figure}

Figure \ref{Fig:3} describes the behavior of the energy conditions. Particulary, Figures. \ref{Fig:3} \subref{fig:WEC}, \subref{fig:NEC}, \subref{fig:SEC} and \subref{fig:DEC} means that all of the energy conditions, WEC,  NEC,  SEC and DEC ones, are satisfied.
\begin{figure}
\centering
\subfigure[~Radial EoS ]{\label{fig:EoS}\includegraphics[scale=0.3]{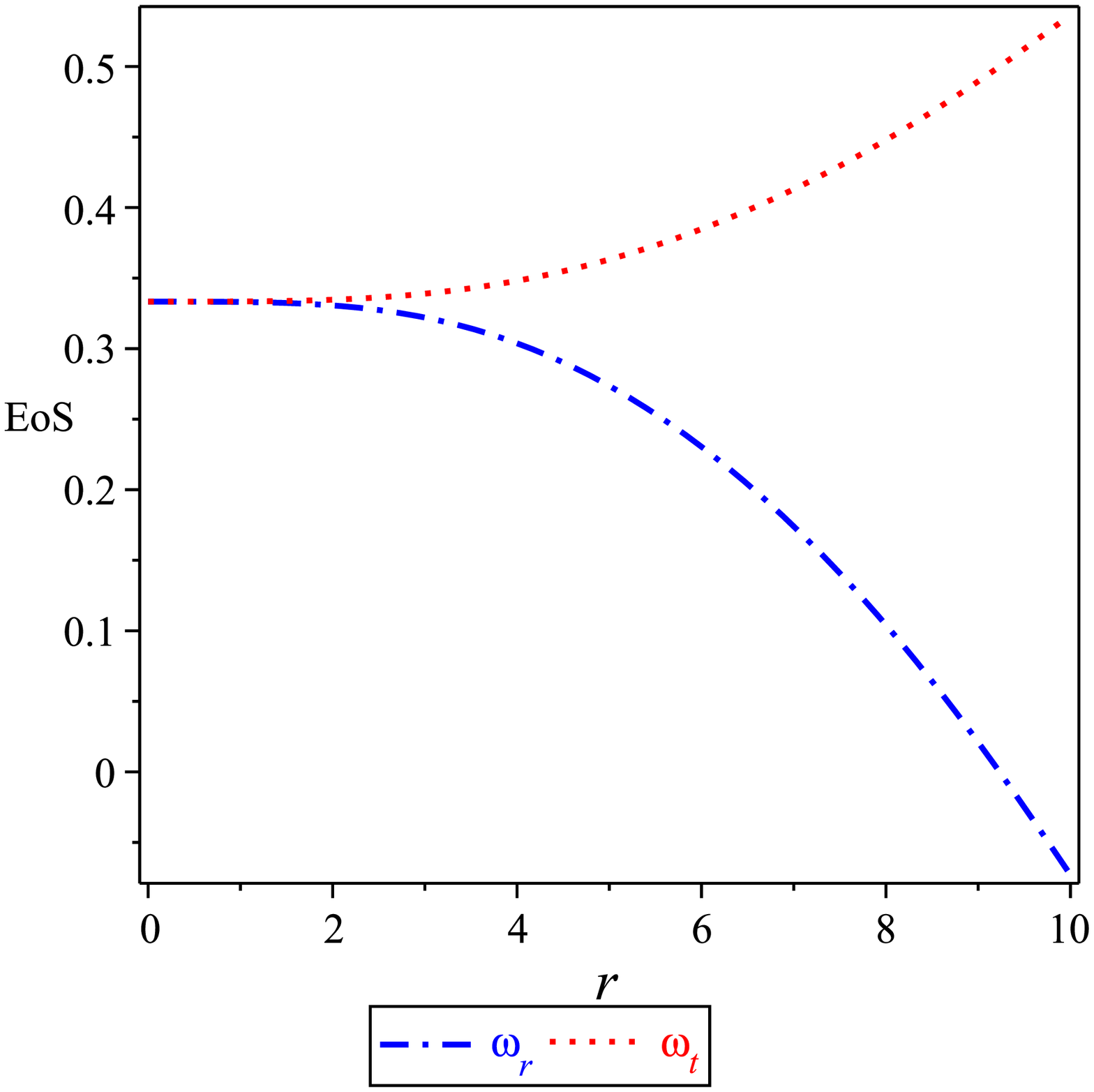}}
\subfigure[~Mass]{\label{fig:mass}\includegraphics[scale=.3]{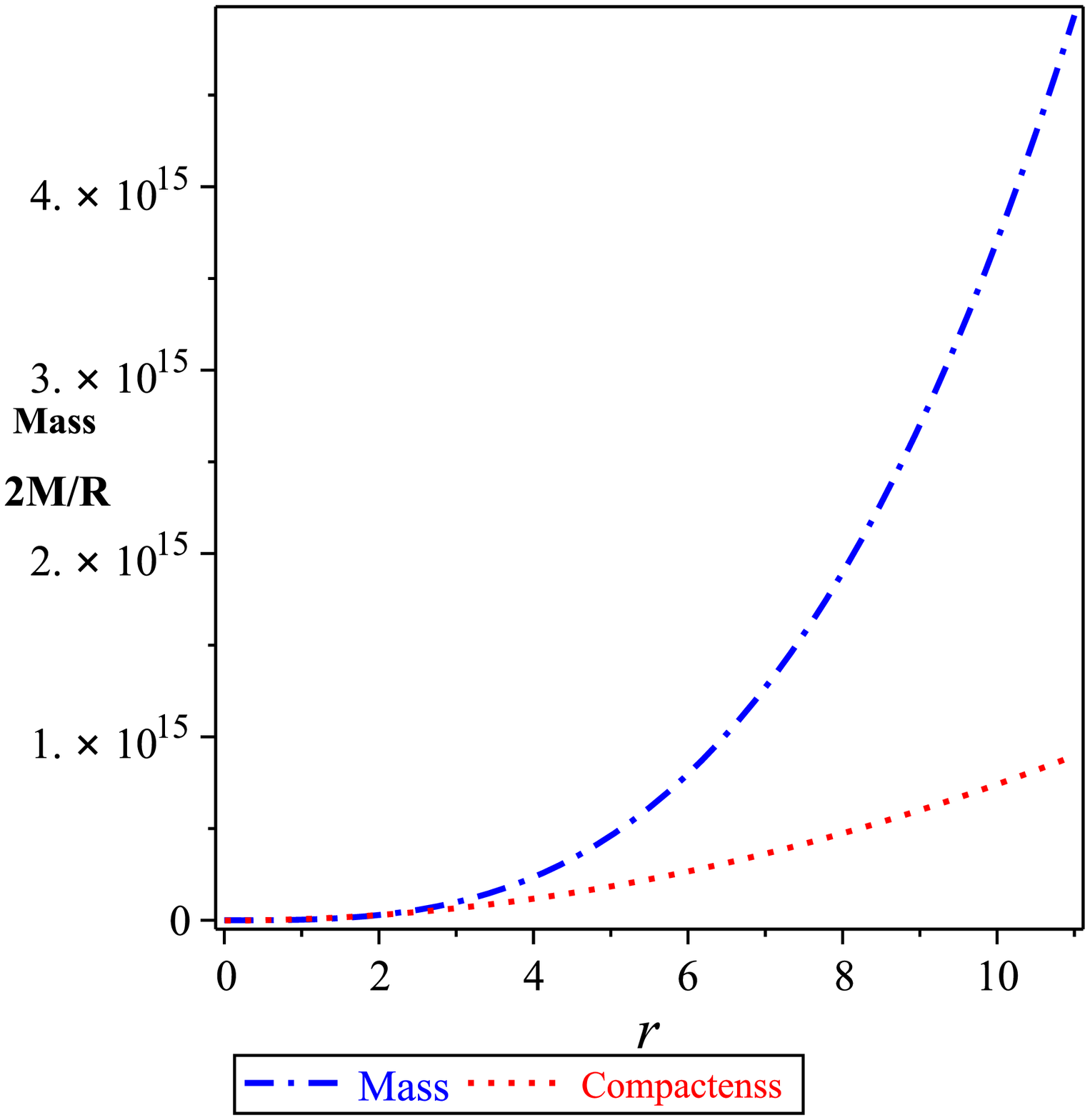}}
\subfigure[~Red-shift]{\label{fig:red}\includegraphics[scale=.3]{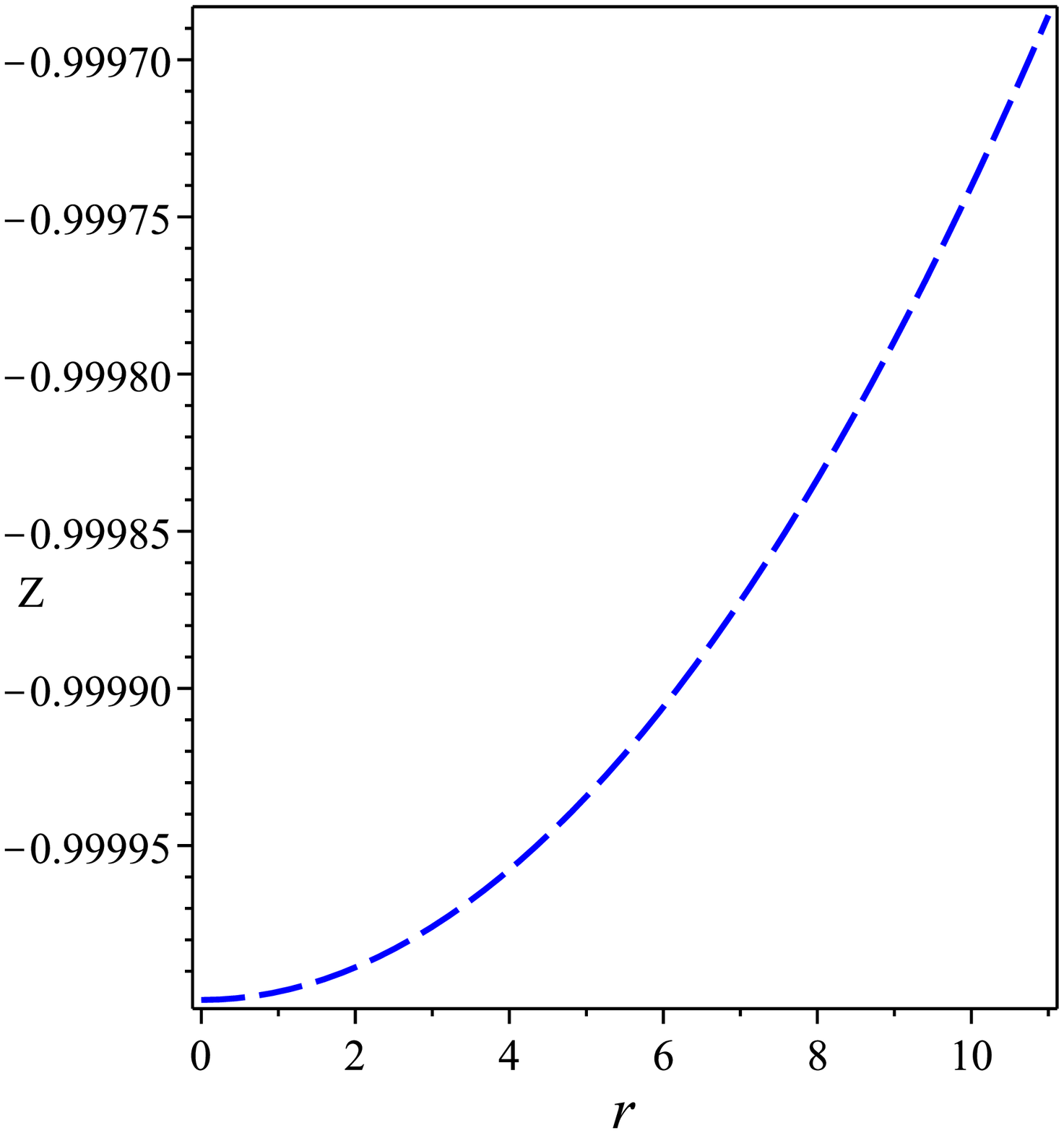}}
\caption[figtopcap]{\small{{Plot of ~\subref{fig:EoS} EoS, ~\subref{fig:mass} mass and compactness,  and ~\subref{fig:red}  the red shift. All of the figures are depicted  vs. the radial coordinate $r$.}}}
\label{Fig:4}
\end{figure}
In Figure  \ref{Fig:4}, we have depicted the radial and the tangential equation of state (EoS). As Fig. \ref{Fig:4} \subref{fig:EoS} shows, the EoS is not linear contrary to what discussed in Ref. \cite{Das:2019dkn}.
From Fig. \ref{Fig:4} \subref{fig:EoS}, it can be found that
the radial and tangential EoS are non-linear, and that the mass function in Eq. (\ref{mas1}) becomes $M(r\rightarrow 0) = 0$.
Additionally, Fig. \ref{Fig:4} \subref{fig:mass} suggests that the compactness parameter becomes larger. Moreover, the radial variation of the surface red-shift is depicted in  Fig. \ref{Fig:4}  \subref{fig:red}.
The present result of $\approx -1$ for $\textrm{4U 1608-52}$ is compatible with the constraints of the surface red-shift $Z\leq 5$ \cite{Bohmer2006}.

\section{Study of the stability of the model}\label{stability}
It is the aim of this section, to study  the stability problem through three different methods; the Tolman-Oppenheimer-Volkoff equations (TOV), the adiabatic index and the static  state.

\subsection{Equilibrium process using the TOV equation}
We explore a hydrostatic equilibrium with the TOV equation \cite{PhysRev.55.364,PhysRev.55.374,PoncedeLeon1993}
\begin{eqnarray}\label{TOV}   \frac{2[p_t-p_r]}{r}-\frac{M_g(r)[\rho(r)+p_r]\sqrt{ab}}{r}-\frac{dp_r}{r}+\frac{\xi[p_t-p_r]}{r}=0,
 \end{eqnarray}
with $M_g(r)$ the gravitational mass at
radius $r$, given by
\begin{eqnarray}\label{ma}   M_g(r)=4\pi{\int_0}^r\Big({T_t}^t-{T_r}^r-{T_\theta}^\theta-{T_\phi}^\phi\Big)r^2\sqrt{\frac{a}{b}}dr=\frac{ra'}{2a\sqrt{b}}\,,
 \end{eqnarray}
Inserting Eq. (\ref{ma}) into (\ref{TOV}), we find
\begin{eqnarray}\label{ma1}  \frac{2(p_t-p_r)}{r}-\frac{dp_r}{dr}-\frac{a'[\rho(r)+p_r]}{2\sqrt{a}}+\frac{\xi[p_t-p_r]}{r}=F_g+F_a+F_h+F_h+F_s=0\,.
 \end{eqnarray}
Here, $F_g=-\frac{a'[\rho(r)+p_r]}{2\sqrt{a}}$,  $F_a=\frac{2(p_t-p_r)}{r}$, $F_h=-\frac{dp_r}{dr}$ and $F_s=\frac{\xi[p_t-p_r]}{r}$ means the gravitational, anisotropic, hydrostatic and scalar field forces, respectively.
The TOV equation for Eq. (\ref{sol}) is found in Fig. \ref{Fig:5}  \subref{fig:TOV}, which shows the four different forces.
It can ben understood that $F_g, F_a, F_h$ are positive and $F_s$ is dominant,
so that the equilibrium of a star can be maintained.

\subsection{Adiabatic index}
For examine the stability of a star, we investigate the equilibrium state
for a spherically symmetric spacetime.
We analyze the adiabatic index $\Gamma$, given by \cite{1964ApJ...140..417C,1989A&A...221....4M,10.1093/mnras/265.3.533}
\begin{eqnarray}\label{a11}  \Gamma=\left(\frac{\rho+p}{p}\right)\left(\frac{dp}{d\rho}\right)\,,
 \end{eqnarray}
where $\rho$ and $p$ denote the density and pressure of isotropic system.
The equilibrium state is realized for $\Gamma>\frac{4}{3}$ \cite{1975A&A....38...51H}. If $\Gamma=\frac{4}{3}$, the isotropic spacetime is in equilibrium.

For an anisotropic system, the stability of a relativistic sphere is realized for $\Gamma >\gamma$ \cite{10.1093/mnras/265.3.533}. Here,
\begin{eqnarray}\label{ai}  \gamma=\frac{4}{3}-\left\{\frac{4(p_r-p_t)}{3\lvert p'_r\lvert}\right\}_{max}\,.
 \end{eqnarray}
 From  Eqs. (\ref{a11}) and (\ref{ai}),  we get the components of the adiabatic index that are listed in Appendix {\color{blue} (C)}.

In Fig. \ref{Fig:5} \subref{fig:adb}, we depict $\gamma$, $\Gamma_r$ and $\Gamma_t$. It is seen that
$\Gamma_r$ and $\Gamma_t$ are larger than $\gamma$ within a star, and thus
the stability conditions are met.

\subsection{Stability in the static state}
\begin{figure}
\centering
\subfigure[~Different forces ]{\label{fig:TOV}\includegraphics[scale=0.3]{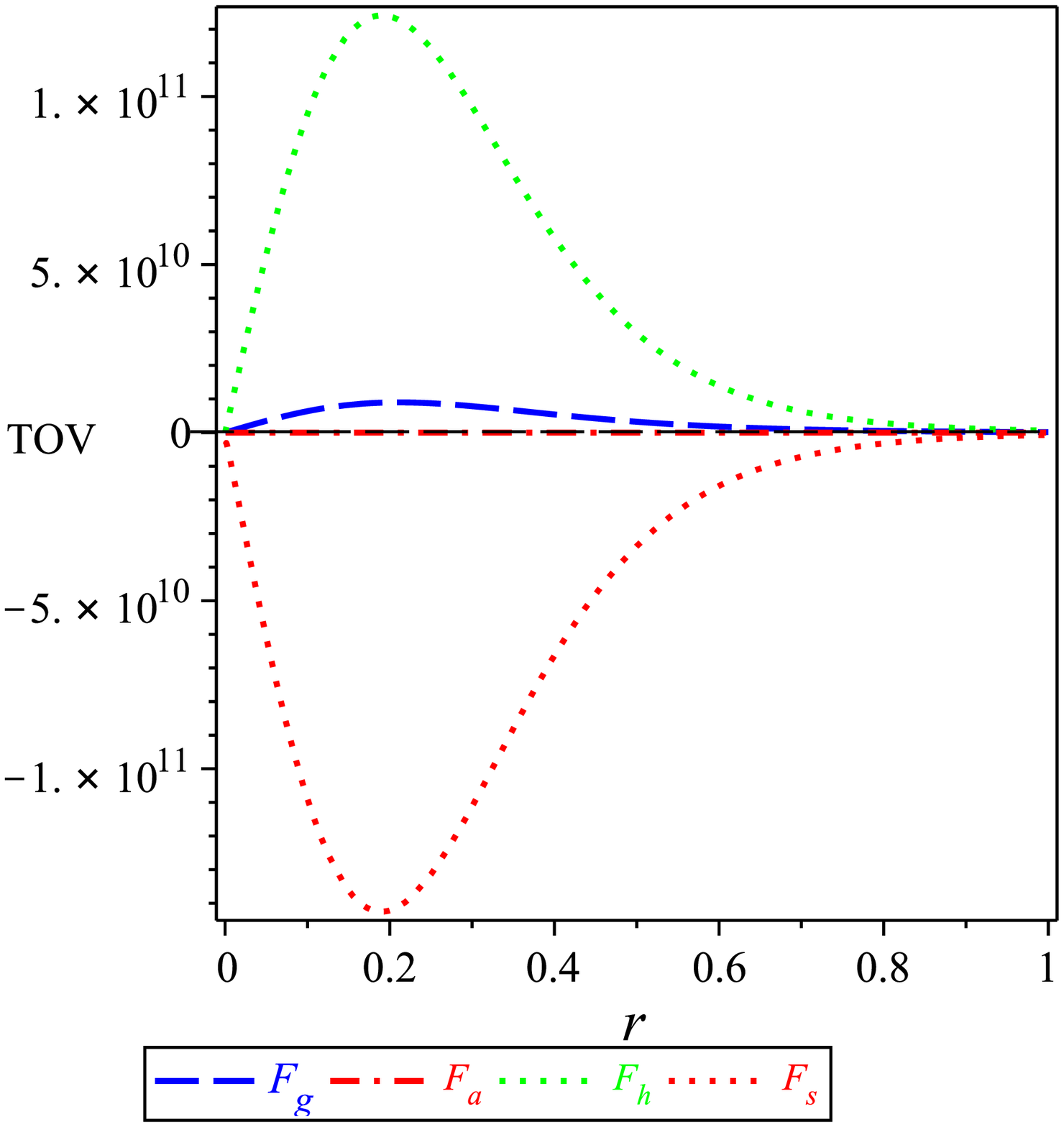}}
\subfigure[~The adiabatic index]{\label{fig:adb}\includegraphics[scale=0.3]{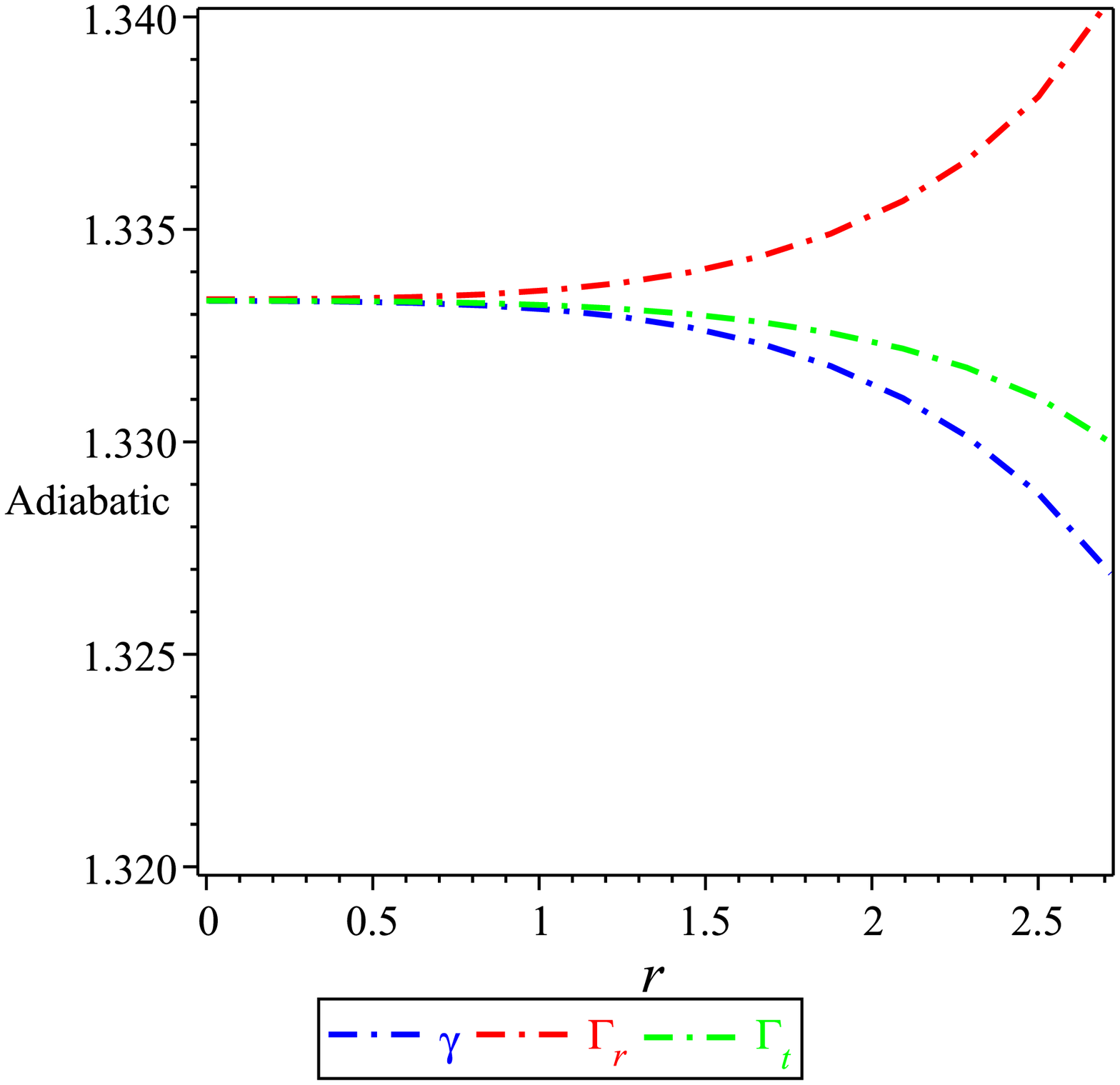}}
\subfigure[~Mass of the central density]{\label{fig:mrho}\includegraphics[scale=0.3]{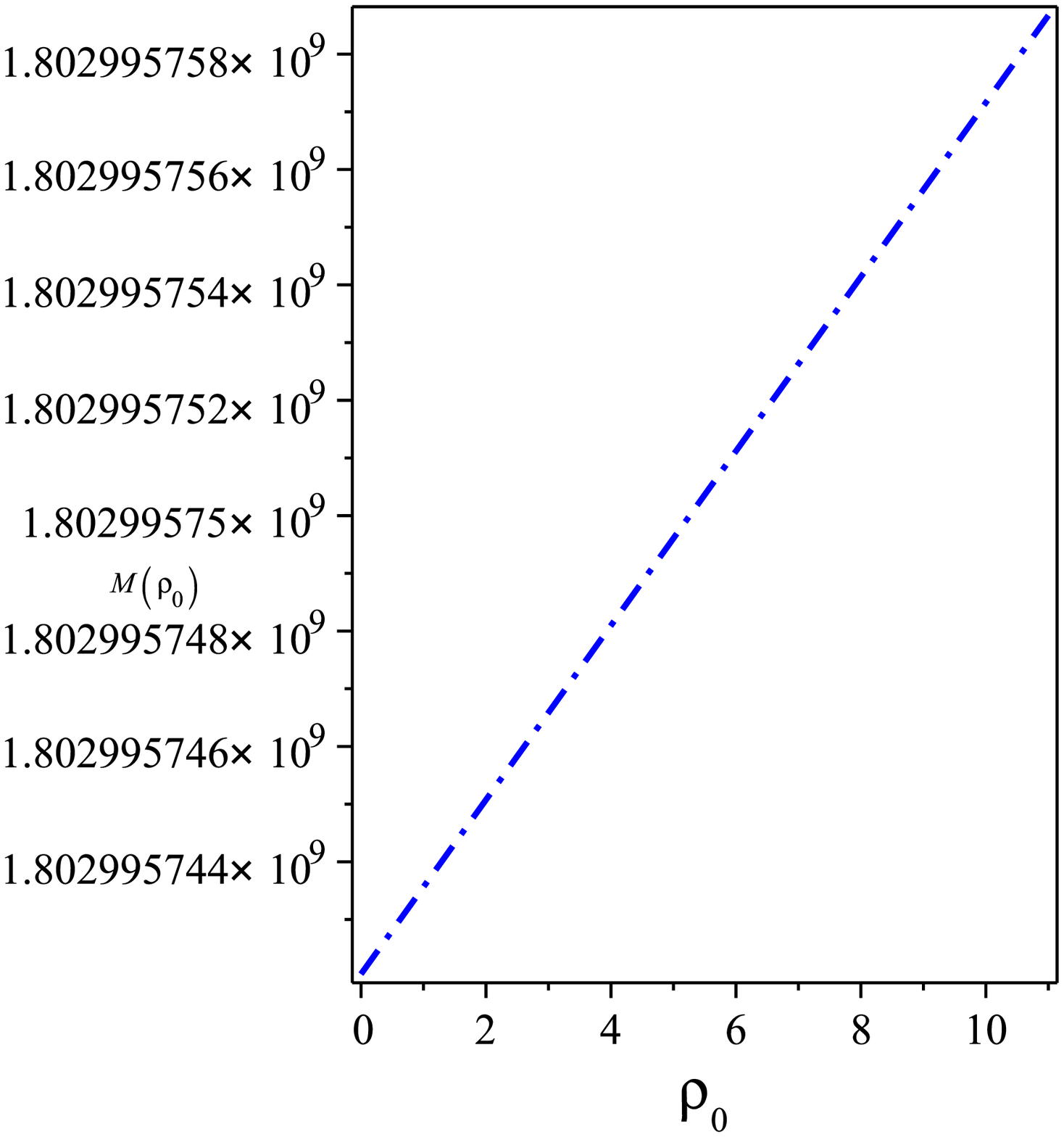}}
\subfigure[~The adiabatic index]{\label{fig:dmrho}\includegraphics[scale=0.3]{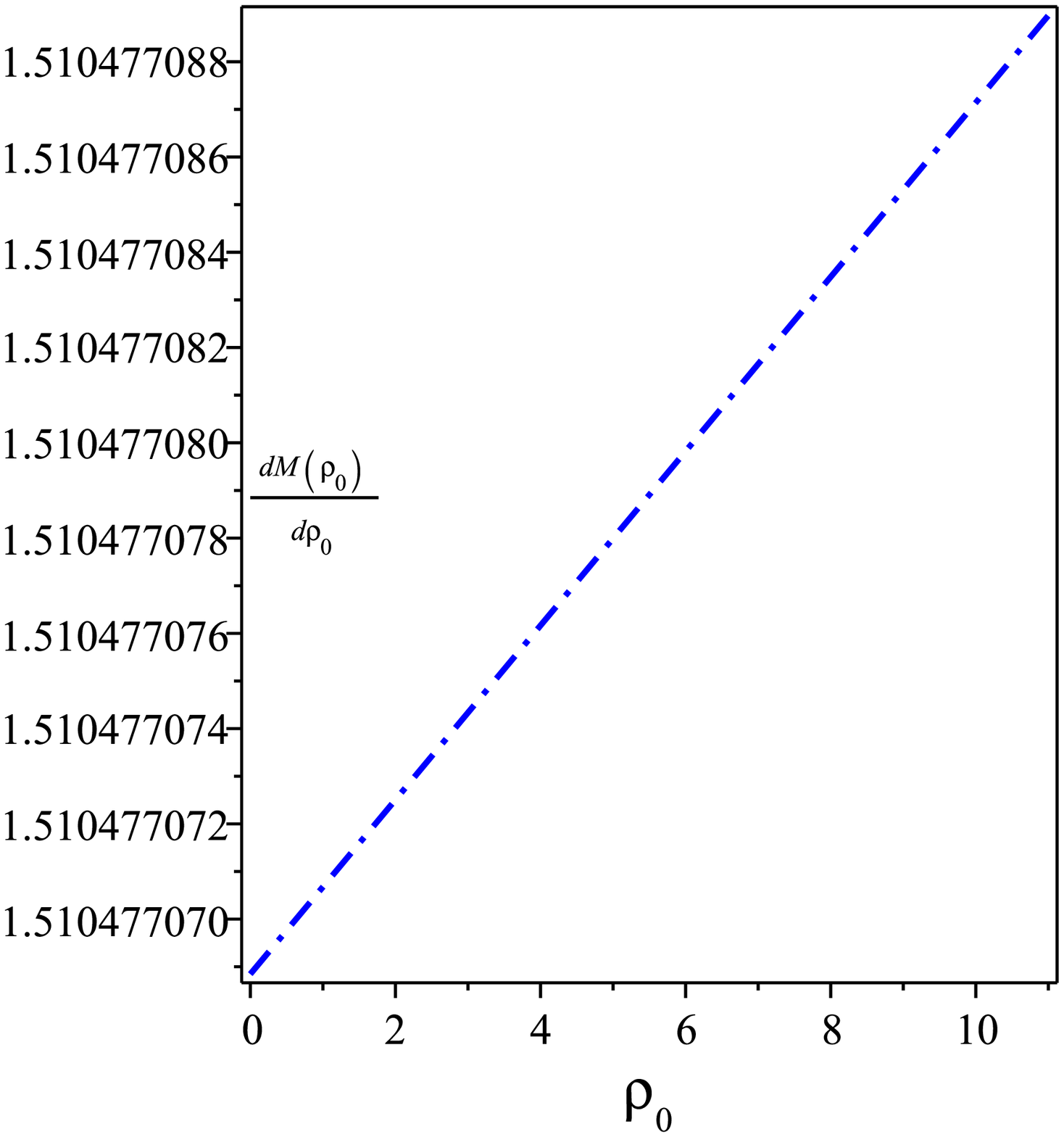}}
\caption[figtopcap]{\small{{Plot of \subref{fig:TOV} the different forces; \subref{fig:adb} the radial and tangential adiabatic indices; \subref{fig:mrho} and \subref{fig:dmrho} the mass and its derivative in terms of the central density. All of the figures are depicted  vs. the radial coordinate $r$.}}}
\label{Fig:5}
\end{figure}

Another test to ensure the stability of the model under consideration is the one of the static state given by Harrison, Zeldovich and Novikov \cite{1965gtgc.book.....H,1971reas.book.....Z,1983reas.book.....Z}. They showed that for a stable star the derivative of the mass in terms of the central density $\rho(\rightarrow 0)$, must give a positive increasing value, i.e.,
$\frac{\partial M}{\partial \rho_0}> 0$. Using this condition in our model we obtain the form of the central density as
\begin{eqnarray}\label{rm}
\rho_0(r\to 0)=\frac{48(a_1c_1-a_0{}^2-a_1a_0{}^2)}{a_1a_0{}^5} \Longrightarrow c_1=\frac{a_0{}^2(\rho_0a_1a_0{}^3+48[1+a_1])}{48a_1}\,.
\end{eqnarray}
Using Eq. (\ref{rm}) in  Eq. (\ref{mas}) we find the mass of the central density, represented as
\begin{eqnarray}
 \label{rm11}
&&M(\rho_0)\approx \,\frac {663552{a_1}^{3}}{{a_0}^{7} \left( \rho_0\,{
 a_1}\,{a_0}^{3}+48+48\,a_1 \right) ^{3}{l}^{5}}+
 \frac{1}{{a_0}^{6} \left( \rho_0
\,a_1\,{a_0}^{3}+48+48\,a_1 \right)^{3}{l}^{3}}\biggl[\,{\frac { 27648(\rho_0\,a_1\,{a_0}^{3}+48+
48\,a_1)}{a_1}}-9216 \biggl( {\frac {288}{{a_1}}}\nonumber\\
&&-{\frac {a_0\, \left( 1344\,{a_0}^{7}{a_1}^{3}-
7{a_0}^{7} \left( \rho_0\,a_1\,{a_0}^{3}+48+48
\,a_1 \right) {a_1}^{3}+192\,{a_0}^{7}{a_1}^{4}
 \right) -288\,{a_1}^{4}{a_0}^{8}}{{a_0}^{8}{a_1}^{4}}} \biggr) \biggr] {a_1}^{3}\,.
\end{eqnarray}
The pattern of the derivative of the mass in terms of the central
density is given by the following asymptotic form
\begin{eqnarray}
 \label{rm111}
&&\frac{\partial M(\rho_0)}{\partial \rho_0}\approx {\frac {1990656{a_1}^{4}}{{a_0}^{4} \left( \rho_0a_1{a_0}^{3}+48+48a_1 \right) ^{4}{l}^{5}}}+
\,{\frac {11520{a_1}^{3}}{{a_0}^{3} \left( \rho_0\,a_1\,{a_0}^{3}+48+48\,a_1 \right) ^{3}{l}^{3}}}-
\frac{{a_1}^{4}}{{a_0}^{3} \left( \rho_0
a_1{a_0}^{3}+48+48\,a_1 \right)^{4}{l}^{3}} \biggl[ {\frac {82944}{a_1}} \biggl(48\nonumber\\
&&\rho_0a_1{a_0}^{3}+
48\,a_1\biggr)-27648\left( {\frac {288}{{a_1}}}-{\frac {a_0\left( 1344\,{a_0}^{7}{a_1}^{3}-
7{a_0}^{7} \left( \rho_0a_1{a_0}^{3}+48+48
a_1 \right) {a_1}^{3}+192{a_0}^{7}{a_1}^{4}
 \right) -288{a_1}^{4}{a_0}^{8}}{{a_0}^{8}{a_1}^{4}}}\right)  \biggr] \,.\nonumber\\
&&
\end{eqnarray}
Equations  (\ref{rm11}) and (\ref{rm111}) are depicted in Figs. \ref{Fig:5} \subref{fig:mrho} and Fig. \ref{Fig:5} \subref{fig:dmrho}  which ensure  the
stability of the model.
\newpage
\begin{table*}[t!]
\caption{\label{Table1}%
Model parameters}
\begin{ruledtabular}
\begin{tabular*}{\textwidth}{lcccccc}
{{Pulsar}}                              & Mass ($M_{\odot}$) &      {Radius (km)} &   ${a_0}$  &    {$a_1$}    & {$c_1$} &\\ \hline
&&&&&&\\
Her X-1                                 &  $0.85\pm 0.15$    &      $8.1\pm0.41$ & -0.004802894320   &    0.3043478261         &-0.04483731442\\
Cen X-3                                &  $1.49\pm 0.08$    &      $9.178\pm0.13$ &  -0.005762649954   &  0.001461108724        & -0.03322916139   \\
RX J 1856 -37                           &  $0.9\pm 0.2$      &      $\simeq 6$   &  -0.01507407407     &  -0.08533333333        & -0.07649056652     \\
4U1608 - 52                        &  $1.74\pm 0.14$      &      $9.52\pm0.15$   &   -0.0061541792     &  -0.1509410548        &-0.02864346182       \\
EXO 1785 - 248                        &  $1.3\pm 0.2$      &      $8.849\pm0.4$   & -0.00561175409     & 0.039896205       & -0.0342249748        \\
4U1820 - 30                        &  $1.58$      &      $9.1$   &  -0.006206183624    & -0.02786813187        &-0.03430457972        \\
&&&&&&\\
\end{tabular*}
\end{ruledtabular}
\end{table*}
\begin{table*}[t!]
\caption{\label{Table1}%
Values of physical quantities}
\begin{ruledtabular}
\begin{tabular*}{\textwidth}{lcccccccccc}
{{Pulsar}}                              &{$\rho\lvert_{_{_{0}}}$} &      {$\rho\lvert_{_{_{l}}}$} &   {$\frac{dp_r}{d\rho}\lvert_{_{_{0}}}$}  &    {$\frac{dp_r}{d\rho}\lvert_{_{_{l}}}$}
  & {$\frac{dp_t}{d\rho}\lvert_{_{_{0}}}$} & {$\frac{dp_t}{d\rho}\lvert_{_{_{l}}}$}&{$(\rho-p_r-2p_t)\lvert_{_{_{0}}}$}&{$(\rho-p_r-2p_t)\lvert_{_{_{l}}}$}&{$z\lvert_{_{_{l}}}$}&\\ \hline
&&&&&&&&&&\\
Her X-1                        &$\approx$8.44$\times10^{11}$     &$\approx$ 0.0012 &$\approx$ 0.3  &$\approx$-2.1    & $\approx$0.3&$\approx$1.2&$\approx$2508.8&$\approx$2.3$\times10^{-14}$& -1  \\
Cen X-3                            &$\approx$4.23$\times10^{11}$     &$\approx$ -0.04 &$\approx$ 0.3  &$\approx$0.7    & $\approx$0.3&$\approx$0.16&$\approx$1.41$\times10^{8}$ &$\approx$1.49$\times10^{-13}$& -1   \\
RX J 1856 -37                   &4.57$\times10^{9}$&-0.052&.33  &3.76 &.33&55.7 &1.5$\times10^{6}$&-0.25$\times10^{-4}$&-1    \\
4U1608 - 52                     & 1.55$\times10^{11}$   &  -0.02051145188 & 0.33  &-2&0.33 &0.36&5.2$\times10^{7}$&-2.34$\times10^{-13}$&-1    \\
EXO 1785 - 248                     & 3$\times10^{11}$     & 0.0105   &.3      &1.57  & .3&.1 &1.48$\times10^{11}$&.0.041&-1\\
4U1820 - 30                             &3$\times10^{11}$ & -0.007  & 0.3 &-0.3 &0.3&1.36 &1.48$\times10^{11}$&.0.06&-1    \\
\end{tabular*}
\end{ruledtabular}
\end{table*}

In addition to PSR J 1614-2230, the same procedure can be performed for other pulsars. The results for other stellar objects are shown in Table I and II

\section{Discussions and conclusions}
A TEGR theory is known to be identical with general relativity \cite{Golovnev_2017,Kr_k_2019,universe5060158}. However, conformal TEGR is not equivalent to the conformal GR theory. In this research, we studied conformal TEGR field equations. In the present paper, we have explored the role of the conformal field on the anisotropic compact objects. Applying the field equations of conformal TEGR theory to a non-diagonal vierbein field. For exterior solution, the scalar torsion is vanishing when the two unknown functions approach Minkowski spacetime which ensures that the vierbein field we begin is a physical one. The output differential equations consist of five non-linear ones in eight unknowns,  $a(r)$, $b(r)$, $\rho(r)$, $p_r(r)$, $p_t(r)$, $V(r)$, $\xi(r)$  and $\Delta(r)$.

As a result, we have obtained an analytic solution for realistic compact star configurations. It has been demonstrated that the model proposed in this work is regular at the center as well as at the surface according to the stellar structure. In particular, the properties of the density as well as the radial and tangential pressures are regular for the case of a pulsar $\textit {4U1608-52}$ \cite{duBuisson:2019djp}.

Moreover, we have shown that the anisotropy and anisotropic forces have negative values which can be considered as attractive forces. This result originates from the fact that the radial pressure is stronger than the tangential pressure, that is, $p_r>p_t $ \cite{SJM19}. The present model is stable owing to the balance of the hydrostatic, gravitational, anisotropic and scalar forces. Also with the procedures of the adiabatic index for a static stat, the model is shown to be stable. Furthermore, we have evaluated the sound speed and confirmed that it is consistent with the observations. According to this result, the metric potentials (\ref{pot}) and the anisotropy expression (\ref{d2}) are considered to be realistic for astrophysical compact objects. In addition, we have analyzed the present model by using the observational data of pulsars (Tables {\color{blue} I} and {\color{blue} II}) and shown that the present theoretical model fit can fit the observational date well.

The consequences obtained in this work are summarized as follows. We have employed a non-diagonal vierbein field yielding the vanishing torsion scalar as the metric potentials are close to unity. Such a behavior is important for the vierbein field as shown in Refs.~\cite{PhysRevD.98.064047,Abbas:2015yma,Momeni:2016oai,2015Ap&SS.359...57A,Chanda:2019hyh,Debnath:2019yge,Ilijic:2018ulf}. It has also been demonstrated that the theoretical predictions of the present model are compatible with the real observational data of pulsars. It is expected that the present study will be able to be tested for a wide range of the metric and anisotropy.


\section*{Acknowledgement}
The work of KB was supported in part by the JSPS KAKENHI Grant Number JP21K03547.

\appendix
{
\section{The gradients of the components of Eqs. (\ref{sol}) \label{A}}
The explicate form of the gradients of the components of the energy-momentum components given by Eq. (\ref{sol}) have the following form:
 \begin{eqnarray}\label{dsol}
&&\rho'= 256\,\Bigg\{ \Bigg[20 \,{a_0}^{6}{r}^{4} -24\,{a_0}^{4}-24\,{r}^{2}{a_0}^{5}-64\,a_1\,{a_0}^{4}-6\,{c_1}^{ 2} \mathrm{S}^{2}-64\,{ a_1}\,c_1\,\mathrm{S} {a_0}^ {2}-52\,{r}^{4}{a_0}^{2}{c_1}^{2}\mathrm{S} -6\,{r}^{2}a_0\,{c_1}^{2}\mathrm{S}^{2}-24\,{r}^{2}{a_0}^ {3}c_1\,\mathrm{S} \nonumber\\
 &&-24\,{r}^{2}a_0\,{c_1}^{2}\mathrm{S}+5\,{a_0}^{2}{r}^{4}{c_1}^{2}\mathrm{S}^{2}+128\,{r}^{2}{a_0}^{3}c_1\,a_1+ 20\,{a_0}^{4}{r}^{4}c_1\,\mathrm{S} -104\,{r}^{2}a_1\,{a_0}^{3}c_1\,\mathrm{S} +64\,{r}^{2}a_0\,{c_1}^{2}a_1\,\mathrm{S} -26\,{r}^{2}a_1\,a_0\,{c_1}^{2} \mathrm{S}^{2}\nonumber\\
 &&+72\,c_1\,{a_1}^{2}{a_0}^{2}+24\,\mathrm{S} {c_1}^{2}{r}^{4}{a_0}^{2 }a_1+32\,\mathrm{S} {c_1}^{2}{  a_0}\,{r}^{2}{a_1}^{2}-40\,\mathrm{S} c_1\,{r}^{4}{a_0}^{4}a_1-20\,\mathrm{S} c_1\,{a_0}^{3}{r}^{2}{a_1}^{2} -10\, \mathrm{S}^{2}{c_1}^{2}{r}^{4}{a_0}^{2}a_1-5\, \mathrm{S}^{2}{c_1}^{2}a_0\,{r}^{2}{{  a_1}}^{2}\nonumber\\
 &&-60\,{c_1}^{2}{a_1}^{2}-20\,{a_1}^{2}{{  a_0}}^{4}+20\,{r}^{6}{a_0}^{7}+48\,c_1\,{r}^{4}{{ a_0 }}^{4}a_1+64\,c_1\,{a_0}^{3}{r}^{2}{a_1}^{2}-60 \,{c_1}^{2}a_0\,{r}^{2}{a_1}^{2}-20\,\mathrm{S}c_1\,{a_1}^{2}{a_0}^{2}+5\,  \mathrm{S}^{2}{c_1 }^{2}{r}^{6}{a_0}^{3}\nonumber\\
 &&-32\,\mathrm{S} { c_1}^{2}{r}^{6}{a_0}^{3}+20\,\mathrm{S} c_1\,{r}^{6}{a_0}^{5}+60\,{c_1}^{2}{r}^{6} {a_0}^{3}-64\,c_1\,{r}^{6}{a_0}^{5}-40\,{r}^{4}{a_0}^{6}a_1-20\,{a_0}^{5}{r}^{2}{a_1}^{2}+36\,\mathrm{S} {c_1}^{2}{a_1}^{2}-5\,  \mathrm{S}^{2}{c_1 }^{2}{a_1}^{2}\nonumber\\
 &&+96\,{a_0}^{2}c_1\,a_1-104\,{r}^{2 }a_1\,{a_0}^{5}-16\,a_1\,{c_1}^{2}\mathrm{S} ^{2}-24\,c_1\,\mathrm{S}{a_0}^{2}-104\,{r}^{4}{a_0 }^{4}c_1+60\,{a_0}^{2}{r}^{4}{c_1}^{2}+48\,{c_1}^{2}a_1\,\mathrm{S} -48\,{r}^{2} {a_0}^{3}c_1 \Bigg] {a_0}^{6}r\Bigg\} \nonumber\\
 &&\Bigg\{ \mathrm{S} \Bigg[ c_1\,\mathrm{S} {r}^{2}a_0-a_1\,c_1\,\mathrm{S} -2\,a_1\,{a_0}^{2} +2\,{a_0}^{3}{r}^{2} \Bigg]\Bigg(   \mathrm{S}^{5}{c_1}^{5}{r}^{2}a_0- \mathrm{S}^{5}{c_1}^{5}a_1-10\,\mathrm{S}^{4}{c_1 }^{4}a_1\,{a_0}^{2}+10\, \mathrm{S}^{4}{c_1}^{4}{r}^{2}{a_0}^{3}+40\,  \mathrm{S}^{3}{c_1 }^{3}{r}^{2}{a_0}^{5}\nonumber\\
 &&-40\, \mathrm{S} ^{3}{c_1}^{3}a_1\,{a_0}^{4}+80\,\mathrm{S}^{2}{c_1 }^{2}{r}^{2}{a_0}^{7}-80\,  \mathrm{S}^{2}{c_1}^{2}a_1\,{a_0}^{6}+80\, \mathrm{S} c_1\,{r}^{2}{a_0}^{9 }-80\,\mathrm{S} c_1\,{a_0}^{8} a_1-32\,a_1\,{a_0}^{10}+32\,{r}^{2}{a_0}^{11} \Bigg)\Bigg\}^{-1} \,, \nonumber\\
&& p'_r=-256\Bigg\{ \Bigg[ 12\,{a_0}^{5}{r}^{2}{a_1}^{2}-4\,{r}^{ 4}{a_0}^{2}{c_1}^{2}\mathrm{S}+2\,{r}^{2}a_0\,{c_1}^{2} \mathrm{S} ^{2}+8\,{r}^{2}{a_0}^{3}c_1\,\mathrm{S}+8\,{r}^{2}a_0\,{c_1}^{2 }\mathrm{S}+5\,{a_0}^{2}{r}^{4}{c_1}^{2} \mathrm{S}^{2}+ 20\,{a_0}^{4}{r}^{4}c_1\,\mathrm{S}\nonumber\\
&& +48\,c_1\,{r}^{4}{a_0}^{4}a_1-32\,c_1 \,{a_0}^{3}{r}^{2}{a_1}^{2}+20\,{c_1}^{2}a_0\,{r }^{2}{a_1}^{2}+12\,\mathrm{S}c_1\,{a_1}^{2}{a_0}^{2}+5\, \mathrm{S}  ^{2}{c_1}^{2}{r}^{6}{a_0}^{3}-16 \,\mathrm{S}{c_1}^{2}{r}^{6}{a_0}^{3}+20\,\mathrm{S}c_1\,{r}^{6} {a_0}^{5}\nonumber\\
&&-40\,{r}^{2}a_1\,{a_0}^{3}c_1\,\mathrm{S} -10\,{r}^{2}a_1\,a_0\,{c_1}^{2}\mathrm{S}^ {2}+24\,\mathrm{S}{c_1}^{2}{r}^{4}{ a_0}^{2}a_1-16\,\mathrm{S}{c_1}^{2}a_0\,{r}^{2}{a_1}^{2}-40\,\mathrm{S} c_1\,{r}^{4}{a_0}^{4}a_1\nonumber\\
&&+12\,\mathrm{S}c_1\,{a_0}^{3}{r}^{2}{a_1}^{2}-10\,\mathrm{S}^{2}{c_1}^{2}{r}^{4}{a_0}^{2}a_1+3\,\mathrm{S}^{2}{c_1}^{2}a_0\,{r}^{2}
{a_1}^{2}+8\,{r}^{2}{a_0}^{5}+20\,{a_0} ^{6}{r}^{4}+2\,{c_1}^{2} \mathrm{S}^{2}+20\,{c_1}^{2}{a_1}^{2}+12\,{a_1}^{2}{a_0}^{4}+20\,{r}^{6}{a_0}^{7}\nonumber\\
&&+3\, \mathrm{S}^{2}{c_1}^{2}{a_1}^{2}- 40\,{r}^{4}{a_0}^{6}a_1-12\,\mathrm{S} {c_1}^{2}{a_1}^{2}-32\,{a_0}^{2}c_1\, a_1-32\,c_1\,{r}^{6}{a_0}^{5}-20\,{c_1}^{2}{r} ^{6}{a_0}^{3}-24\,c_1\,{a_1}^{2}{a_0}^{2}-40\,{r }^{2}a_1\,{a_0}^{5}+8\,c_1\,\mathrm{S} {a_0}^{2}\nonumber\\
&&-8\,{r}^{4}{a_0}^{4}c_1-20\, {a_0}^{2}{r}^{4}{c_1}^{2}-16\,{c_1}^{2}a_1\, \mathrm{S}+16\,{r}^{2}{a_0}^{3}c_1+8\,{a_0}^{4} \Bigg] {a_0}^{6}r\Bigg\}\Bigg\{ \left( 1+{r}^{2}a_0 \right)  \left( c_1\,\mathrm{S} {r}^{2}a_0-a_1\,c_1\,\mathrm{S} -2\,a_1\,{a_0}^{2}+2\,{a_0}^{3}{r}^{2} \right)  \nonumber\\
&&\Bigg[   \mathrm{S}^{5}{c_1}^{5}{r}^{2}a_0- \mathrm{S}^{5}{c_1}^{5}a_1-10\,\mathrm{S}^{4}{c_1 }^{4}a_1\,{a_0}^{2}+10\, \mathrm{S}^{4}{c_1}^{4}{r}^{2}{a_0}^{3}+40\,\mathrm{S}^{3}{c_1 }^{3}{r}^{2}{a_0}^{5}-40\,\mathrm{S}^{3}{c_1}^{3}a_1\,{a_0}^{4}+80\, \mathrm{S}^{2}{c_1 }^{2}{r}^{2}{a_0}^{7}-80\, \mathrm{S}^{2}{c_1}^{2}a_1\,{a_0}^{6}\nonumber\\
&&+80\, \mathrm{S}c_1\,{r}^{2}{a_0}^{9 }-80\,\mathrm{S}c_1\,{a_0}^{8} a_1-32\,a_1\,{a_0}^{10}+32\,{r}^{2}{a_0}^{11} \Bigg] \Bigg\}^{-1} \nonumber\\
&&p'_t=-256\,\Bigg\{ \Bigg[ 32\,{r}^{2}a_0\,{c_1}^{2}a_1\, \mathrm{S} -4\,{a_0}^{5}{r}^{2}{a_1}^{2}-28\,{r}^{4}{a_0}^{2}{c_1}^{2}\mathrm{S}-2\,{r}^{2}a_0\,{c_1}^{2} \mathrm{S}^{2}-8\,{r}^{2}{a_0}^{3}c_1\,\mathrm{S} -8\,{r}^{2}a_0\,{c_1}^{2}\mathrm{S} +5\,{a_0}^{2}{r}^{4}{c_1}^{2} \mathrm{S}^{2}\nonumber\\
&&+20\,{a_0}^{4}{r}^{4}c_1\,\mathrm{S} +48\,c_1\,{r}^{4}{a_0}^{ 4}a_1+16\,c_1\,{a_0}^{3}{r}^{2}{a_1}^{2}-20\,{c_1}^{2}a_0\,{r}^{2}{a_1}^{2}-4\,\mathrm{S} c_1\,{a_1}^{2}{a_0}^{2}+5\,\mathrm{S}^{2}{c_1}^{2}{r} ^{6}{a_0}^{3}-24\,\mathrm{S} {c_1}^{2}{r}^{6}{a_0}^{3}+20\,\mathrm{S} c_1\,{r}^{6}{a_0}^{5}\nonumber\\
&&-32\,a_1\,c_1\, \mathrm{S} {a_0}^{2}-72\,{r}^{2}a_1\,{a_0}^{3}c_1\,\mathrm{S} - 18\,{r}^{2}a_1\,a_0\,{c_1}^{2}\mathrm{S}^{2}+24\,\mathrm{S} {c_1}^{2}{r}^{4}{a_0}^{2}a_1+8\,\mathrm{S} {c_1}^{2}a_0\,{r}^{2}{a_1}^ {2}-40\,\mathrm{S} c_1\,{r}^{4}{a_0}^{4}a_1-4\,\mathrm{S} c_1\,{a_0}^{3}{r}^{2}{a_1}^{2}\nonumber\\
&&-10\,\mathrm{S}^{2}{c_1}^{2}{r}^{4}{a_0}^ {2}a_1- \mathrm{S}^{2 }{c_1}^{2}a_0\,{r}^{2}{a_1}^{2}-8\,{r}^{2}{a_0}^ {5}+20\,{a_0}^{6}{r}^{4}-2\,{c_1}^{2} \mathrm{S}^{2}-20\,{c_1}^{2}{a_1}^{2}-4\,{a_1}^{2}{a_0}^{4}+20\,{r}^{6}{a_0}^{7}-\mathrm{S}^{2}{c_1}^{2}{a_1}^{2}-40\,{r}^{4}{a_0}^{6}a_1\nonumber\\
&&+12\,\mathrm{S} {c_1}^{2}{a_1}^{2}+32\,{a_0}^{2}c_1\,a_1-48\,c_1\,{r}^{6}{a_0}^{5}+20\,{c_1}^{2}{r}^{6}{a_0}^{3}+24\,c_1\,{a_1}^{2}{a_0}^{2}-72\,{r}^{2} a_1\,{a_0}^{5}-8\,c_1\,\mathrm{S} {a_0}^{2}-56\,{r}^{4}{a_0}^{4}c_1+20\,{a_0}^{2}{r}^{4}{c_1}^{2}\nonumber\\
&&+16\,{c_1}^{ 2}a_1\,\mathrm{S} -16\,{r}^{2}{a_0}^{3}c_1-8\,{a_0}^{4}+64\,{r}^{2}a_1\,{a_0}^{3} c_1-8\,{c_1}^{2} \mathrm{S}^{2}a_1-32\,a_1\,{a_0}^{4} \Bigg] {a_0}^{6}r\Bigg\}\Bigg\{ \left( 1+{r}^{2}a_0 \right)  \Bigg[ c_1\, \mathrm{S} {r}^{2}a_0-a_1\,c_1\,\mathrm{S} -2\,a_1\,{a_0}^ {2}\nonumber\\
&&+2\,{a_0}^{3}{r}^{2} \Bigg]  \Bigg( \mathrm{S}^{5}{c_1}^{5}{r}^{2}a_0- \mathrm{S}^{5}{c_1 }^{5}a_1-10\, \mathrm{S}^{4}{c_1}^{4}a_1\,{a_0}^{2}+10\,\mathrm{S}^{4}{c_1}^{4}{r}^{2} {a_0}^{3}+40\, \mathrm{S} ^{3}{c_1}^{3}{r}^{2}{a_0}^{5}-40\, \mathrm{S}^{3}{c_1}^{3}a_1\,{a_0}^{4}+80\,\mathrm{S}^{2}{c_1}^{2}{r}^{2}{a_0}^{7}\nonumber\\
&&-80\,\mathrm{S}^{2}{c_1}^{2}a_1\,{a_0}^{6}+80\,\mathrm{S} c_1 \,{r}^{2}{a_0}^{9}-80\,\mathrm{S} c_1\,{a_0}^{8}a_1-32\,a_1\,{a_0}^{10}+32\,{r }^{2}{a_0}^{11} \Bigg) \Bigg\}^{-1} \, ,
\end{eqnarray}
where $\rho'=\frac{d\rho}{dr}$, $p'_r=\frac{dp_r}{dr}$ and $p'_t=\frac{dp_t}{dr}$. Eqs. (\ref{dsol}) show that the gradients of density, radial and tangential pressures are negative as we will show  when we plot them.\vspace{0.1cm}\\
\section{The components of the speed of sound  \label{B}}
 Using Eq. (\ref{sol}) we evaluate the value of the components of the sound and get:
 \begin{eqnarray}\label{dso2}
&&v_r{}^2=-\biggl\{20\,{c_1 }^{2}{a_1}^{2}-24\,c_1\,{a_1}^{2}{a_0}^{2}+12\,{a_1}^{2}{a_0}^{4}+20\,{r}^{6}{a_0}^{7}+8\,c_1\,\mathrm{S} a_0{}^{2}+16\,{r}^{2}{a_0}^{3}c_1-32\,{a_0}^{2}c_1\,a_1-8\,{r}^{4}{a_0}^{4}c_1-16\,{c_1}^{2}a_1\,\mathrm{S}\nonumber\\
&&-20\,{a_0}^{2}{r}^ {4}{c_1}^{2}-40\,{r}^{2}a_1\,{a_0}^{5}+8\,{r}^{2}{a_0}^{5}+20\,{a_0}^{6}{r}^{4}+2\,{c_1}^{2} \mathrm{S}^{2}-12\,\mathrm{S} {c_1}^{2}{a_1}^{2}+20\,{c_1}^{2 }a_0\,{r}^{2}{a_1}^{2}-32\,c_1\,{a_0}^{3}{r}^{2} {a_1}^{2}+48\,c_1\,{r}^{4}{a_0}^{4}a_1\nonumber\\
&&+12\,\mathrm{S} c_1\,{a_1}^{2}{a_0}^{2}+5\,\mathrm{S}^{2}{c_1}^{2}{r}^{6}{a_0}^{3}-16\,\mathrm{S}{c_1}^{2}{r}^{6}{a_0}^{3}+20\,\mathrm{S} c_1\,{r}^{6}{a_0}^{5}-32\,c_1\,{ r}^{6}{a_0}^{5}-40\,{r}^{4}{a_0}^{6}a_1+12\,{a_0}^ {5}{r}^{2}{a_1}^{2}-20\,{c_1}^{2}{r}^{6}{a_0}^{3}\nonumber\\
&&+3\,\mathrm{S}^{2}{c_1 }^{2}{a_1}^{2}+3\,\mathrm{S}^{2}{c_1}^{2}a_0\,{r}^{2}{a_1}^{2}+24\,\mathrm{S} {c_1}^{2}{r}^{4}{a_0}^{2 }a_1-16\,\mathrm{S}{c_1}^{2}a_0\,{r}^{2}{a_1}^{2}+12\,\mathrm{S}c_1\,{a_0}^{3}{r}^{2}{a_1}^{2}-40\,\mathrm{S} c_1\,{r}^{4}{a_0}^{4}a_1-10\,\mathrm{S}^{2 }{c_1}^{2}{r}^{4}{a_0}^{2}a_1\nonumber\\
&&+8\,{a_0}^{4}-40\,{ r}^{2}a_1\,{a_0}^{3}c_1\,\mathrm{S} -10\,{r}^{2}a_1\,a_0\,{c_1}^{2} \mathrm{S}^{2}+20\,{a_0}^{4}{r}^ {4}c_1\,\mathrm{S}-4\,{r}^{4}{a_0}^{2}{c_1}^{2}\mathrm{S}+2\,{r}^ {2}a_0\,{c_1}^{2}\mathrm{S}^{2}+8\,{r}^{2}{a_0}^{3}c_1\,\mathrm{S} +8\,{r}^{2}a_0\,{c_1}^{2 }\mathrm{S}\nonumber\\
&&+5\,{a_0}^{2}{r}^{4}{c_1}^{2}\mathrm{S}^{2}\biggr\} \times\biggl\{72\,c_1\,{a_1}^{2}{a_0}^{2}-60\,{c_1}^{2}{a_1}^{2}-20\,{a_1}^{2}{a_0}^{4}+20\,{r}^{6}{a_0}^{7 }-24\,c_1\,\mathrm{S} {a_0}^{2} -48\,{r}^{2}{a_0}^{3}c_1+96\,{a_0}^{2}c_1\,a_1-104\,{r}^{4}{a_0}^{4}c_1\nonumber\\
&&+4\,{c_1}^{2}[12a_1\,\mathrm{S}+15\,{a_0}^{2}{r}^{4}-4\,a_1 \mathrm{S}^{2}]-104\,{r}^{2}a_1\,{a_0}^{ 5}-24\,{r}^{2}{a_0}^{5}+20\,{a_0}^{6}{r}^{4}-64\,a_1\,{ a_0}^{4}-6\,{c_1}^{2}\mathrm{S}^{2}+36\,\mathrm{S} {c_1}^{2}{a_1}^{2}-60\,{c_1}^{2}a_0\,{r}^{2}{a_1}^{2}\nonumber\\
&&+64\,c_1\,{a_0}^{3}{r}^{2}{a_1}^{2}+48\,c_1\,{r}^{4}{a_0}^{4}a_1-20\,\mathrm{S}c_1\,{a_1}^{2}{a_0}^{2}+5\,\mathrm{S}^{2}{c_1}^{2}{r} ^{6}{a_0}^{3}-32\,\mathrm{S}{c_1}^{2}{r}^{6}{a_0}^{3}+20\,\mathrm{S}c_1\,{r}^{6}{a_0}^{5}-64\,c_1\,{r}^{6}{a_0}^{5}-40\,{r}^{4}{a_0}^{6}a_1\nonumber\\
&&-20\,{a_0}^{5}{r}^{ 2}{a_1}^{2}+60\,{c_1}^{2}{r}^{6}{a_0}^{3}-5\,\mathrm{S}^{2}{c_1}^{2}{a_1}^{2}-5\, \mathrm{S}^{2}{c_1}^{2}a_0\,{r}^{2}{a_1}^{2}+24\,\mathrm{S} {c_1}^{2}{r}^{4}{a_0}^{2 }a_1+32\,\mathrm{S} {c_1}^{2}a_0\,{r}^{2}{a_1}^{2}-20\,\mathrm{S}c_1\,{a_0}^{3}{r}^{2}{a_1}^{2}\nonumber\\
&&-40\,\mathrm{S} c_1\,{r}^{4}{a_0}^{4}a_1-10\,\mathrm{S}^{2 }{c_1}^{2}{r}^{4}{a_0}^{2}a_1-24\,{a_0}^{4}-104 \,{r}^{2}a_1\,{a_0}^{3}c_1\,\mathrm{S}+64\,{r}^{2}a_0\,{c_1}^{2}a_1\,\mathrm{S} -26\,{r}^{2}a_1\,a_0\,{c_1}^{2}\mathrm{S}^ {2}+20\,{a_0}^{4}{r}^{4}c_1\,\mathrm{S}\nonumber\\
&&-64\,a_1\,c_1\,\mathrm{S} {a_0}^{2}-52\,{r}^{4}{a_0}^{2}{c_1}^{2}\mathrm{S} -6\,{r}^{2}a_0\,{c_1}^{2 }\mathrm{S}^{2}-24\,{r}^{ 2}{a_0}^{3}c_1\,\mathrm{S}-24 \,{r}^{2}a_0\,{c_1}^{2}\mathrm{S} +5\,{a_0}^{2}{r}^{4}{c_1}^{2}\mathrm{S}^{2}+128\,{r}^{2}{a_0}^{3}c_1\,a_1\biggr\}^{-1} , \nonumber\\
&&v_r{}^2=\biggl\{24\,c_1\,{a_1}^{2}{a_0}^{2}-20\,{c_1}^ {2}{a_1}^{2}-4\,{a_1}^{2}{a_0}^{4}+20\,{r}^{6}{a_0 }^{7}-8\,c_1\,\mathrm{S} {a_0}^ {2}-16\,{r}^{2}{a_0}^{3}c_1+32\,{a_0}^{2}c_1\, a_1-56\,{r}^{4}{a_0}^{4}c_1+16\,{c_1}^{2}a_1\,\mathrm{S}\nonumber\\
&& +20\,{a_0}^{2}{r}^{4}{ c_1}^{2}-8\,a_1\,{c_1}^{2} \mathrm{S}^{2}-72\,{r}^{2}a_1\,{a_0}^{5} -8\,{r}^{2}{a_0}^{5}+20\,{a_0}^{6}{r}^{4}-32\,a_1\,{a_0}^{4}-2\,{c_1}^{2} \mathrm{S}^{2}+12\,\mathrm{S} {c_1}^{2}{a_1}^{2}-20\,{c_1}^{2}a_0\,{r}^{2}{a_1}^{2}\nonumber\\
&&+16\,c_1\,{a_0}^{3}{r}^{2}{a_1}^{2}+48c_1\,{r}^{4}{a_0}^{4}a_1-4\,\mathrm{S} c_1\,{a_1}^{2}{a_0}^{2}+5\, \mathrm{S}^{2}{c_1}^{2}{r}^{6} {a_0}^{3}-24\,\mathrm{S} {c_1}^ {2}{r}^{6}{a_0}^{3}+20\,\mathrm{S}c_1\,{r}^{6}{a_0}^{5}-48\,c_1\,{r}^{6}{a_0}^{5} -40\,{r}^{4}{a_0}^{6}a_1\nonumber\\
&&-4\,{a_0}^{5}{r}^{2}{a_1}^ {2}+20\,{c_1}^{2}{r}^{6}{a_0}^{3}- \mathrm{S}^{2}{c_1}^{2}{a_1}^{2}-\mathrm{S}^{2}{c_1 }^{2}a_0\,{r}^{2}{a_1}^{2}+24\,\mathrm{S} {c_1}^{2}{r}^{4}{a_0}^{2}a_1+8\,\mathrm{S} {c_1}^{2}a_0\,{r}^{2}{a_1}^ {2}-4\,\mathrm{S} c_1\,{a_0}^{3 }{r}^{2}{a_1}^{2}-40\,\mathrm{S} c_1\,{r}^{4}{a_0}^{4}a_1\nonumber\\
&&-10\, \mathrm{S}^{2}{c_1}^{2}{r}^{4}{a_0}^{2}a_1-8\,{a_0}^{4}-72\,{r}^{2}a_1\,{a_0}^{3}c_1\,\mathrm{S} +32\,{r}^{2}a_0\,{c_1}^{2}a_1\,\mathrm{S} -18\,{r}^{2} a_1\,a_0\,{c_1}^{2}\mathrm{S}^{2}+20\,{a_0}^{4}{r}^{4}c_1\,\mathrm{S} -32\,a_1\,c_1\,\mathrm{S} {a_0}^{2}-28\,{r}^{4}{a_0} ^{2}{c_1}^{2}\mathrm{S} \nonumber\\
&&-2\,{r}^{2}a_0\,{c_1}^{2} \mathrm{S} ^{2}-8\,{r}^{2}{a_0}^{3}c_1\,\mathrm{S} -8\,{r}^{2}a_0\,{c_1}^{2 }\mathrm{S} +5\,{a_0}^{2}{r}^{4}{c_1}^{2} \mathrm{S}^{2}+ 64\,{r}^{2}{a_0}^{3}c_1\,a_1\biggr\}\biggl\{72\,c_1\,{a_1}^{2}{a_0}^{2}-60\,{c_1}^{2}{a_1}^{2}-20\,{a_1}^{2}{a_0}^{4}+20\,{r}^{6}{a_0}^{7}\nonumber\\
&&-24\,c_1\,\mathrm{S} {a_0}^{2}-48\,{r}^{2}{a_0} ^{3}c_1+96\,{a_0}^{2}c_1\,a_1-104\,{r}^{4}{a_0}^{4}c_1+48\,{c_1}^{2}a_1\,\mathrm{S} +60\,{a_0}^{2}{r}^{4}{c_1}^{2}-16\,a_1\,{c_1}^{2} \mathrm{S}^{2}-104\,{r}^{2}a_1\,{a_0}^{5}-24\,{r}^{2 }{a_0}^{5}\nonumber\\
&&+20\,{a_0}^{6}{r}^{4}-64\,a_1\,{a_0}^{4} -6\,{c_1}^{2} \mathrm{S}^{2}+36\,\mathrm{S} {c_1}^{ 2}{a_1}^{2}-60\,{c_1}^{2}a_0\,{r}^{2}{a_1}^{2}+ 64\,c_1\,{a_0}^{3}{r}^{2}{a_1}^{2}+48\,c_1\,{r }^{4}{a_0}^{4}a_1-20\,\mathrm{S} c_1\,{a_1}^{2}{a_0}^{2}+5\, \mathrm{S} ^{2}{c_1}^{2}{r}^{6}{a_0}^{3}\nonumber\\
&& -32\,\mathrm{S} {c_1}^{2}{r}^{6}{a_0}^{3}+20\,\mathrm{S} c_1\,{r} ^{6}{a_0}^{5}-64\,c_1\,{r}^{6}{a_0}^{5}-40\,{r}^{4}{a_0}^{6}a_1-20\,{a_0}^{5}{r}^{2}{a_1}^{2}+60\,{c_1}^{2}{r}^{6}{a_0}^{3}-5\, \mathrm{S}^{2}{c_1}^{2}{a_1}^{2}-5\, \mathrm{S}^{2}{c_1 }^{2}a_0\,{r}^{2}{a_1}^{2}\nonumber\\
&&+24\,\mathrm{S} {c_1}^{2}{r}^{4}{a_0}^{2}a_1+32\,\mathrm{S} {c_1}^{2}a_0\,{r}^{2}{a_1}^{2}-20\,\mathrm{S} c_1\,{a_0}^{3}{r}^{2}{a_1}^{2}-40\,\mathrm{S} c_1\,{r}^{4}{a_0}^{4}a_1-10\, \mathrm{S}^{2}{c_1}^{2}{r}^{4} {a_0}^{2}a_1-24\,{a_0}^{4}-104\,{r}^{2}a_1\,{a_0}^{3}c_1\,\mathrm{S}\nonumber\\
&& +64\,{r}^{2} a_0\,{c_1}^{2}a_1\,\mathrm{S} -26\,{r}^{2}a_1\,a_0\,{c_1}^{2} \mathrm{S}^{2}+20\,{a_0}^{4}{r}^ {4}c_1\,\mathrm{S} -64\,a_1\,c_1\,\mathrm{S} {a_0}^{2}-52\,{r }^{4}{a_0}^{2}{c_1}^{2}\mathrm{S} -6\,{r}^{2}a_0\,{c_1}^{2} \mathrm{S}^{2}-24\,{r}^{2}{a_0}^{3}c_1\,\mathrm{S} -24\,{r}^{2}a_0\,{c_1}^{2}\mathrm{S} \nonumber\\
&&+5\,{a_0}^{2}{r}^ {4}{c_1}^{2}\mathrm{S}^{2}+128\,{r}^{2}{a_0}^{3}c_1\,a_1\biggr\}
\, .\nonumber\\
 &&
\end{eqnarray}
\section{The components of the adiabatic index \label{C}}
The explicate form of the  components of the adiabatic index  given by Eq. (\ref{sol}) take  the following form:
 \begin{eqnarray}\label{a12}
&&\Gamma=-8\biggl\{16\,c_1\,{a_0}^{3}{r}^{2}{a_1}^{2}-4\, \mathrm{S}c_1\,{a_1}^{2}{a_0}^{2}+8\,\mathrm{S}{c_1}^{2}{r}^ {6}{a_0}^{3}-8\,a_1\,c_1\,\mathrm{S}{a_0}^{2}-4\,c_1\,\mathrm{S} {a_0}^{2}-8\,{r}^{2}{a_0}^{3}c_1+16\,{a_0}^{2}c_1\,a_1+4\,{r}^{4}{a_0}^{4}c_1\nonumber\\
&&+8\,{c_1}^{2}a_1\,\mathrm{S}+10\,{a_0}^{2}{r}^{4}{c_1}^{2}+12\,c_1\,{a_1}^{2}{a_0}^{2}+6\,\mathrm{S}{c_1}^{2} {a_1}^{2}+10\,{c_1}^{2}{r}^{6}{a_0}^{3}+16\,c_1\,{r}^{6}{a_0}^{5}+8\,{r}^{4}{a_0}^{6}a_1-4\,{a_0 }^{5}{r}^{2}{a_1}^{2}- \mathrm{S} ^{2}{c_1}^{2}{a_1}^{2}\nonumber\\
&&-2\,a_1\,{c_1}^{2}\mathrm{S}^ {2}+4\,{r}^{2}{a_0}^{5}+8\,{a_0}^{6}{r}^{4}-{c_1}^{2}\mathrm{S} ^{2}-10\,{c_1}^{2}{a_1}^{2}-4\,{a_1}^{2}{a_0}^{4}-8\,a_1\, {a_0}^{4}-4\,{a_0}^{4}+8\,\mathrm{S}c_1\,{r}^{4}{a_0}^{4}a_1-4\,\mathrm{S}c_1\,{a_0}^{3}{r}^{2}{a_1}^{2}\nonumber\\
&&+ 2\, \mathrm{S} ^{2}{c_1}^{2}{r}^{4}{a_0}^{2}a_1-12\,\mathrm{S} {c_1}^{2}{r}^{4}{a_0}^{2}a_1+8\mathrm{S} {c_1}^{2}a_0\,{r}^{2}{a_1}^{2}-\mathrm{S}^{ 2}{c_1}^{2}a_0\,{r}^{2}{a_1}^{2}+8\,{a_0}^{4}{r} ^{4}c_1\,\mathrm{S}+2\,{r}^{4}{a_0}^{2}{c_1}^{2}\mathrm{S}+{r} ^{2}a_0\,{c_1}^{2} \mathrm{S}^{2}+4\,{r}^{2}{a_0}^{3}c_1\,\mathrm{S}\nonumber\\
&& -4\,{r}^{2}a_0\,{c_1}^{2 }\mathrm{S}+2\,{a_0}^{2}{r}^{4}{c_1}^{2}\mathrm{S}^{2}- 10\,{c_1}^{2}a_0\,{r}^{2}{a_1}^{2}-24\,c_1\,{r }^{4}{a_0}^{4}a_1\biggr\}\biggl\{3\biggl(-32\,c_1\,{a_0}^{3}{r}^{2}{a_1}^{2}+12\,\mathrm{S}c_1\,{a_1}^{2}{a_0}^{2}+5\, \mathrm{S}^{2}{c_1}^{2}{r}^{6}{a_0}^{3}\nonumber\\
&&-16\,\mathrm{S}{c_1}^{2}{r}^{6}{a_0}^{3 }+20\,\mathrm{S} c_1\,{r}^{6}{a_0}^{5}+8\,c_1\,\mathrm{S}{{a_0 }}^{2}+16\,{r}^{2}{a_0}^{3}c_1-32\,{a_0}^{2}c_1\,a_1-8\,{r}^{4}{a_0}^{4}c_1-16\,{c_1}^{2}{ a_1}\,\mathrm{S}-20\,{a_0}^{2}{r}^ {4}{c_1}^{2}-40\,{r}^{2}a_1\,{a_0}^{5}\nonumber\\
&&-24\,c_1 \,{a_1}^{2}{a_0}^{2}-12\,\mathrm{S} {c_1}^{2}{a_1}^{2}-20\,{c_1}^{2}{r}^{6}{a_0}^{3}-32\,c_1\,{r}^{6}{a_0}^{5}-40\,{r}^{4}{{a_0 }}^{6}a_1+12\,{a_0}^{5}{r}^{2}{a_1}^{2}+3\,\mathrm{S}^{2}{c_1}^{2}{a_1}^{2}+8\,{r}^{2}{a_0}^{5}+20\,{a_0}^{6}{r}^{4}\nonumber\\
&&+2\,{c_1}^{2}\mathrm{S}^{2}+ 20\,{c_1}^{2}{a_1}^{2}+12\,{a_1}^{2}{a_0}^{4}+20 \,{r}^{6}{a_0}^{7}+8\,{a_0}^{4}-40\,{r}^{2}a_1\,{a_0}^{3}c_1\,\mathrm{S}-10\,{r}^{2} a_1\,a_0\,{c_1}^{2} \mathrm{S}^{2}-40\,\mathrm{S}c_1\,{r}^{4}{a_0}^{4}a_1+12\,\mathrm{S}\,c_1\,{a_0}^{3}{r}^{2}{a_1}^{2}\nonumber\\
&&-10\,\mathrm{S}^{2}{c_1 }^{2}{r}^{4}{a_0}^{2}a_1+24\,\mathrm{S}{c_1}^{2}{r}^{4}{a_0}^{2}a_1-16\,\mathrm{S}{c_1}^{2}a_0\,{r}^{2}{a_1}^{2}+3\,\mathrm{S}^{2}{c_1}^{2}a_0
\,{r}^{2}{a_1}^{2}+20\,{a_0}^{4}{r}^{4}c_1\,\mathrm{S}- 4\,{r}^{4}{a_0}^{2}{c_1}^{2}\mathrm{S}+2\,{r}^{2}a_0\,{c_1}^{2} \mathrm{S}^{2}\nonumber\\
&&+8\,{r}^{2}{a_0}^{3}c_1 \,\mathrm{S}+8\,{r}^{2}a_0\,{c_1}^{2}\mathrm{S}+5\,{a_0}^{2}{r}^ {4}{c_1}^{2}\mathrm{S}^{2}+20\,{c_1}^{2}a_0\,{r}^{2}{a_1}^{2}+48\,
c_1\,{r}^{4}{a_0}^{4}a_1\biggr)\biggr\}\,.
 \end{eqnarray}

 From Eq. (\ref{a11}),  we obtain  the adiabatic index of solution (\ref{sol})   in the form
 \begin{eqnarray}\label{aic}
&&\Gamma_r=8\,\biggl\{ \biggl[ -2\,c_1\,{r}^{4}{a_0}^{2}-2\,c_1 \,{r}^{2}a_0+2\,{r}^{2}a_1\,{a_0}^{3}-2\,c_1\,{ a_1}+2\,{a_0}^{3}{r}^{2}+2\,a_1\,{a_0}^{2}+2\,{a_0}^{2}-2\,c_1\,a_0\,{r}^{2}a_1+a_1\,c_1 \,\mathrm{S} +c_1\,\mathrm{S} {r}^{2}a_0+c_1\,\mathrm{S}\nonumber\\
&& +{r}^{2}a_1\,a_0\,c_1\,\mathrm{S} \biggr]\times  \biggl( -32\,c_1\,{a_0}^ {3}{r}^{2}{a_1}^{2}+12\,\mathrm{S} c_1\,{a_1}^{2}{a_0}^{2}+5\, \mathrm{S}^{2}{c_1}^{2}{r}^{6}{a_0}^{3}- 16\,\mathrm{S} {c_1}^{2}{r}^{6}{a_0}^{3}+20\,\mathrm{S} c_1\,{r} ^{6}{a_0}^{5}+8\,c_1\,\mathrm{S} {a_0}^{2}+16\,{r}^{2}{a_0}^{3}c_1\nonumber\\
&& -32\,{a_0}^{2}c_1\,a_1-8\,{r}^{4}{a_0}^{4}c_1-16\,{{ \it \_C1}}^{2}a_1\,\mathrm{S} -20\,{{ a_0}}^{2}{r}^{4}{c_1}^{2}-40\,{r}^{2}a_1\,{a_0}^{5 }-24\,c_1\,{a_1}^{2}{a_0}^{2}-12\,\mathrm{S} {c_1}^{2}{a_1}^{2}-20\,{c_1}^{2} {r}^{6}{a_0}^{3}-32\,c_1\,{r}^{6}{a_0}^{5}\nonumber\\
&& -40\,{r}^{4 }{a_0}^{6}a_1+12\,{a_0}^{5}{r}^{2}{a_1}^{2}+3\,\mathrm{S}^{2}{c_1 }^{2}{a_1}^{2}+8\,{r}^{2}{a_0}^{5}+20\,{a_0}^{6}{r}^{4} +2\,{c_1}^{2} \left( \mathrm{S} \right) ^{2}+20\,{c_1}^{2}{a_1}^{2}+12\,{a_1}^{2}{a_0}^{4}+20\,{r}^{6}{a_0}^{7}+8\,{a_0}^{4}\nonumber\\
&& -40\,{r}^{2}{  a_1}\,{a_0}^{3}c_1\,\mathrm{S} -10\,{r}^{2}a_1\,a_0\,{c_1}^{2} \mathrm{S}^{2}-40\,\mathrm{S}c_1\,{r}^{4}{a_0}^{4}a_1+12\,\mathrm{S}c_1\,{a_0}^{3}{r}^{2}{{ a_1}}^{2}-10\,\mathrm{S}^{2}{c_1}^{2}{r}^{4}{a_0}^{2}a_1+24\,\mathrm{S} {c_1}^{2}{r}^{4}{a_0}^{2 }a_1-16\,\mathrm{S} {c_1}^{2}{  a_0}\,{r}^{2}{a_1}^{2}\nonumber\\
&& +3\, \mathrm{S}^{2}{c_1}^{2}a_0\,{r}^{2}{a_1}^{2} +20\,{a_0}^{4}{r}^{4}c_1\,\mathrm{S} -4\,{r}^{4}{a_0}^{2}{c_1}^{2}\ln  \left( 1+{r}^{2 }a_0 \right) +2\,{r}^{2}a_0\,{c_1}^{2}\mathrm{S}^{2}+8\,{r}^{2}{a_0}^{ 3}c_1\,\mathrm{S} +8\,{r}^{2}a_0\,{c_1}^{2}\mathrm{S} +5\,{a_0 }^{2}{r}^{4}{c_1}^{2}\mathrm{S}^{2}\nonumber\\
&&+20\,{c_1}^{2}a_0\,{r}^{2}{a_1} ^{2}+48\,c_1\,{r}^{4}{a_0}^{4}a_1 \biggr)\biggr\}\times \biggl\{ \biggl[ 2 \,c_1\,\mathrm{S} {r}^{2}a_0+5 \,{a_0}^{2}{r}^{4}c_1\,\mathrm{S} -2\,c_1\,\mathrm{S} -3\,{r} ^{2}a_1\,a_0\,c_1\,\mathrm{S} -2\,a_1\,c_1\,\mathrm{S} +4\,c_1\,a_0\,{r}^{2}a_1+4\,c_1\,a_1\nonumber\\
&&+4\,c_1\,{r}^{2}a_0+4\,c_1\,{r}^{4}{a_0 }^{2}+4\,{a_0}^{3}{r}^{2}-6\,{r}^{2}a_1\,{a_0}^{3}-4\,{ a_0}^{2}+10\,{a_0}^{4}{r}^{4}-4\,a_1\,{a_0}^{2} \biggr]\times  \biggl( 64\,c_1\,{a_0}^{3}{r}^{2}{a_1}^{2}- 20\,\mathrm{S} c_1\,{a_1}^{2}{a_0}^{2}+5\, \mathrm{S} ^{2}{c_1}^{2}{r}^{6}{a_0}^{3}\nonumber\\
&&-32\,\mathrm{S} {c_1}^{2}{r}^{6}{a_0}^{3}+20\,\mathrm{S} c_1\,{r}^{6}{a_0}^{5}-64 \,a_1\,c_1\,\mathrm{S} {{a_0 }}^{2}+128\,{r}^{2}{a_0}^{3}c_1\,a_1-24\,c_1\, \mathrm{S} {a_0}^{2}-48\,{r}^{2}{a_0}^{3}c_1+96\,{a_0}^{2}c_1\,a_1-104\,{r}^{4} {a_0}^{4}c_1\nonumber\\
&&+48\,{c_1}^{2}a_1\,\mathrm{S} +60\,{a_0}^{2}{r}^{4}{c_1}^{2}-104 \,{r}^{2}a_1\,{a_0}^{5}+72\,c_1\,{a_1}^{2}{a_0}^{2}+36\,\mathrm{S} {c_1}^{2}{{ a_1}}^{2}+60\,{c_1}^{2}{r}^{6}{a_0}^{3}-64\,c_1 \,{r}^{6}{a_0}^{5}-40\,{r}^{4}{a_0}^{6}a_1-20\,{{ a_0 }}^{5}{r}^{2}{a_1}^{2}\nonumber\\
&&-5\,\mathrm{S}^{2}{c_1}^{2}{a_1}^{2}-16\,a_1\,{c_!}^{2} \left( \mathrm{S}  \right) ^ {2}-24\,{r}^{2}{a_0}^{5}+20\,{a_0}^{6}{r}^{4}-6\,{c_1 }^{2}\mathrm{S}^{2}-60\,{ c_1}^{2}{a_1}^{2}-20\,{a_1}^{2}{a_0}^{4}+20\,{r} ^{6}{a_0}^{7}-64\,a_1\,{a_0}^{4}-24\,{a_0}^{4}\nonumber\\
&&-104 \,{r}^{2}a_1\,{a_0}^{3}c_1\,\mathrm{S}-26\,{r}^{2}a_1\,a_0\,{c_1}^{2} \mathrm{S}^{2}-40\,\mathrm{S}c_1\,{r}^{4}{a_0}^{4}{ a_1}-20\,\mathrm{S} c_1\,{{a_0 }}^{3}{r}^{2}{a_1}^{2}-10\,\mathrm{S}^{2}{c_1}^{2}{r}^{4}{a_0}^{2}a_1+24 \,\mathrm{S} {c_1}^{2}{r}^{4}{a_0}^{2}a_1\nonumber\\
&&+32\,\mathrm{S} {c_1 }^{2}a_0\,{r}^{2}{a_1}^{2}-5\, \mathrm{S}^{2}{c_1}^{2}a_0\,{r}^{2}{a_1}^{2}+64\,{r}^{2}a_0\,{c_1}^{2}a_1\,\mathrm{S} +20\,{a_0}^{4}{r}^{4}c_1\,\mathrm{S} -52\,{r}^{4}{a_0}^{2}{c_1}^{2}\mathrm{S} -6\,{r}^{2}a_0\,{c_1}^{2} \mathrm{S}^ {2}-24\,{r}^{2}{a_0}^{3}c_1\,\mathrm{S}\nonumber\\
&& -24\,{r}^{2}a_0\,{c_1}^{2}\mathrm{S} +5\,{a_0}^{2}{r}^{4}{c_1}^{2} \mathrm{S}^{2}-60\,{c_1}^{2}a_0\,{r}^{2}{a_1}^{2}+48\,c_1\,{r}^{4}{a_0}^{4}a_1 \biggr) \biggr\}^{-1} \,, \nonumber\\
&&\Gamma_t=2\biggl\{ \biggl(6\,{r}^{2}a_1\,{a_0}^{3} -8\,c_1\,{r}^{4}{a_0}^{2}-8\,c_1 \,{r}^{2}a_0-8\,c_1\,a_1+16\,{a_0}^{3}{r}^{2}+8\,a_1\,{a_0}^{2}+10\,{a_0}^{4}{r}^{4}+8\,{a_0}^{2}-8\,c_1\,a_0\,{r}^{2} a_1+8\,c_1\,\mathrm{S} {r}^{2}a_0\nonumber\\
&& +5\,{a_0}^{2}{r}^{4}c_1\,\mathrm{S} +3\,{r}^{2}a_1\,a_0\,c_1\,\mathrm{S} +4\,c_1\,\mathrm{S} +4\,a_1\,c_1\,\mathrm{S}  \biggr)  \biggl[ 16\,c_1\,{a_0}^{3}{r}^{2}{a_1}^{2}-4\,\mathrm{S} c_1\,{a_1}^{2}{a_0}^{2}+5\,\mathrm{S}^{2}{c_1}^{2}{r}^{6}{a_0}^{3}-24\,\mathrm{S} {c_1}^{2}{r}^{6}{a_0}^{3}+20\,\mathrm{S} c_1\,{r}^{6}{a_0}^{5}\nonumber\\
&& -32 \,a_1\,c_1\,\mathrm{S} {a_0}^{2}+64\,{r}^{2}{a_0}^{3}c_1\,a_1-8\,c_1\, \mathrm{S} {a_0}^{2}-16\,{r}^{2}{a_0}^{3}c_1+32\,{a_0}^{2}c_1\,a_1-56\,{r}^{4}{ a_0}^{4}c_1+16\,{c_1}^{2}a_1\,\mathrm{S} +20\,{a_0}^{2}{r}^{4}{c_1}^{2}-72\,{ r}^{2}a_1\,{a_0}^{5}\nonumber\\
&& +24\,c_1\,{a_1}^{2}{a_0 }^{2}+12\,\mathrm{S} {c_1}^{2}{a_1}^{2}+20\,{c_1}^{2}{r}^{6}{a_0}^{3}-48\,c_1\,{r} ^{6}{a_0}^{5}-40\,{r}^{4}{a_0}^{6}a_1-4\,{a_0}^{5} {r}^{2}{a_1}^{2}- \mathrm{S} ^{2}{c_1}^{2}{a_1}^{2}-8\,a_1\,{c_1}^ {2}\mathrm{S}^{2}-8\,{r}^ {2}{a_0}^{5}\nonumber\\
&& +20\,{a_0}^{6}{r}^{4}-2\,{c_1}^{2} \mathrm{S}^{2}-20\,{c_1}^{2}{a_1}^{2}-4\,{a_1}^{2}{a_0}^{4}+20\,{r}^{6}{a_0}^{7}-32\,a_1\,{a_0}^{4}-8\,{a_0}^{4}-72\,{r}^{2 }a_1\,{a_0}^{3}c_1\,\mathrm{S} -18\,{r}^{2}a_1\,a_0\,{c_1}^{2}\mathrm{S}^{2}\nonumber\\
&& -40\,\mathrm{S} c_1\,{r}^{4}{a_0}^{4}a_1-4\,\mathrm{S} c_1\,{a_0}^{3}{r}^{2}{a_1}^{2}-10\,\mathrm{S} ^{2}{c_1}^{2}{r}^{4}{a_0}^{2}a_1+24\,\mathrm{S} {c_1}^{2}{r}^{4}{a_0}^{2 }a_1+8\,\mathrm{S} {c_1}^{2}a_0\,{r}^{2}{a_1}^{2}-\mathrm{S}^{2}{c_1}^{2}a_0\,{r}^{2}{a_1}^{2}+ 32\,{r}^{2}a_0\,{c_1}^{2}a_1\,\mathrm{S}\nonumber\\
&&  +20\,{a_0}^{4}{r}^{4}c_1\,\mathrm{S} -28\,{r}^{4}{a_0}^{2}{c_1}^{2}\mathrm{S} -2\,{r}^{2}a_0\,{c_1}^{2 } \left( \mathrm{S}  \right) ^{2}-8\,{r}^{2 }{a_0}^{3}c_1\,\mathrm{S} -8\,{ r}^{2}a_0\,{c_1}^{2}\mathrm{S} +5\,{a_0}^{2}{r}^{4}{c_1}^{2} \mathrm{S}^{2}-20\,{c_1}^{2}a_0\,{r}^{2}{a_1}^{2}+48\,c_1\,{r}^{4}{a_0}^{4}a_1 \biggr] \biggr\}\nonumber\\
&& \times\biggl\{ \biggl[ 6\,c_1\,\mathrm{S} {r}^{2}a_0+5\,{a_0}^{2}{r}^{4}c_1\,\mathrm{S} +2\,c_1\,\mathrm{S} +{r} ^{2}a_1\,a_0\,c_1\,\mathrm{S} +2\,a_1\,c_1\,\mathrm{S} -4\,c_1\,a_0\,{r}^{2}a_1-4\,c_1\,a_1-4\,c_1\,{r}^{2}a_0-4\,c_1\,{r}^{4}{a_0 }^{2}+12\,{a_0}^{3}{r}^{2}\nonumber\\
&& +2\,{r}^{2}a_1\,{a_0}^{3}+4\, {a_0}^{2}+10\,{a_0}^{4}{r}^{4}+4\,a_1\,{a_0}^{2} \biggr]  \biggl( 64\,c_1\,{a_0}^{3}{r}^{2}{a_1}^{2}- 20\,\mathrm{S} c_1\,{a_1}^{2}{a_0}^{2}+5\, \mathrm{S}^{2}{c_1}^{2}{r}^{6}{a_0}^{3}-32\,\mathrm{S}{c_1}^{2}{r}^{6}{a_0}^{3}+20\,\mathrm{S} c_1\,{r}^{6}{a_0}^{5}\nonumber\\
&& -64 \,a_1\,c_1\,\mathrm{S} {{ a_0 }}^{2}+128\,{r}^{2}{a_0}^{3}c_1\,a_1-24\,c_1\, \mathrm{S} {a_0}^{2}-48\,{r}^{2}{a_0}^{3}c_1+96\,{a_0}^{2}c_1\,a_1-104\,{r}^{4} {a_0}^{4}c_1+48\,{c_1}^{2}a_1\,\mathrm{S} +60\,{a_0}^{2}{r}^{4}{c_1}^{2}\nonumber\\
&& -104 \,{r}^{2}a_1\,{a_0}^{5}+72\,c_1\,{a_1}^{2}{a_0}^{2}+36\,\mathrm{S} {c_1}^{2}{a_1}^{2}+60\,{c_1}^{2}{r}^{6}{a_0}^{3}-64\,c_1 \,{r}^{6}{a_0}^{5}-40\,{r}^{4}{a_0}^{6}a_1-20\,{a_0}^{5}{r}^{2}{a_1}^{2}-5\, \mathrm{S}^{2}{c_1}^{2}{a_1}^{2}-16\,a_1\,{c_1}^{2} \mathrm{S}^ {2}\nonumber\\
&& -24\,{r}^{2}{a_0}^{5}+20\,{a_0}^{6}{r}^{4}-6\,{c_1 }^{2}\mathrm{S}^{2}-60\,{ c_1}^{2}{a_1}^{2}-20\,{a_1}^{2}{a_0}^{4}+20\,{r} ^{6}{a_0}^{7}-64\,a_1\,{a_0}^{4}-24\,{a_0}^{4}-104 \,{r}^{2}a_1\,{a_0}^{3}c_1\,\mathrm{S} -26\,{r}^{2}a_1\,a_0\,{c_1}^{2}\mathrm{S}^{2}\nonumber\\
&& -40\,\mathrm{S} c_1\,{r}^{4}{a_0}^{4}a_1-20\,\mathrm{S} c_1\,{{ a-0 }}^{3}{r}^{2}{a_1}^{2}-10\, \mathrm{S} ^{2}{c_1}^{2}{r}^{4}{a_0}^{2}a_1+24 \,\mathrm{S} {c_1}^{2}{r}^{4}{a_0}^{2}a_1+32\,\mathrm{S} {c_1 }^{2}a_0\,{r}^{2}{a_1}^{2}-5\, \mathrm{S}^{2}{c_1}^{2}a_0\,{r}^{2}{a_1}^{2}\nonumber\\
&& +64\,{r}^{2}a_0\,{c_1}^{2}a_1\,\mathrm{S} +20\,{a_0}^{4}{r}^{4}c_1\,\mathrm{S} -52\,{r}^{4}{a_0}^{2}{{c_1 }}^{2}\mathrm{S} -6\,{r}^{2}a_0\,{{ c_1}}^{2}\mathrm{S}^ {2}-24\,{r}^{2}{a_0}^{3}c_1\,\mathrm{S} -24\,{r}^{2}a_0\,{c_1}^{2}\mathrm{S} +5\,{a_0}^{2}{r}^{4}{c_1}^{2}\mathrm{S}^{2}\nonumber\\
&& -60\,{c_1}^{2}a_0\,{r}^{2}{a_1}^{2}+48\,c_1\,{r}^{4}{a_0}^{4}a_1 \biggr)\biggr\}^{-1}\,.
 \end{eqnarray}

%

 \end{document}